\documentclass[aps,pre,groupedaddress]{revtex4}

\usepackage{latexsym,amssymb,graphicx,caption}
\usepackage{subfigure}
\begin{document}

\title{Secondary Structures in Long Compact Polymers}
\author{Richard Oberdorf${}^1$}
\author{Allison Ferguson${}^1$}
\author{Jesper L.~Jacobsen${}^{2,3}$}
\author{Jan\'e  Kondev${}^1$}
\affiliation{${}^1$Department of Physics, Brandeis University, Waltham, MA 02454, USA}
\affiliation{${}^2$LPTMS, Universit\'e Paris-Sud, B\^atiment 100, 91405 Orsay, France}
\affiliation{${}^3$Service de Physique Th\'eorique, CEA Saclay, 91191 Gif-sur-Yvette, France}

\date{\today}

\begin{abstract}

Compact polymers are self-avoiding random walks which visit every site on a
lattice. This polymer model is used widely for studying statistical problems
inspired by protein folding. One difficulty with using compact
polymers to perform numerical calculations is generating a
sufficiently large number of randomly sampled configurations.
We present a Monte-Carlo algorithm which uniformly
samples compact polymer configurations in an efficient manner allowing
investigations of chains much longer than previously studied.  Chain
configurations generated by the algorithm are used to compute
statistics of secondary structures in compact polymers. We
determine the fraction of monomers participating in secondary structures,
and show that it is self averaging in the long chain limit and strictly less than one.
Comparison with results for lattice models of open polymer chains shows that compact
chains are significantly more likely to form secondary structure.

\end{abstract}


\maketitle

\section{Introduction}

Proteins are long, flexible chains of amino acids which can assume, in
the presence of a denaturant, an astronomically large number of open
conformations. Twenty different types of amino acids are found in
naturally occurring proteins, and their sequence along the chain
defines the primary structure of the protein. The native, folded state
of the protein contains secondary structures such as $\alpha$-helices
and $\beta$-sheets which are in turn arranged to form the larger tertiary
structures. Under proper solvent conditions most proteins will fold
into a unique native conformation which is determined by its sequence.
One of the goals of protein folding research is to
determine exactly how the folded state results from the specific
sequence of amino acids in the primary structure.

A number of theories exist to describe the forces that are responsible
for protein folding~\cite{dill95}.  Since there are many fewer
compact polymer conformations than non-compact ones, entropic forces resist the tight packing of globular
proteins. Tight packing is primarily the result of
hydrophobic interactions between the amino acid monomers and the solvent molecules around them. Compared to
the local forces between neighboring monomers along the chain, the hydrophobic
interactions were historically seen as nonlocal forces contributing to
the collapse process, but not responsible for determining the
specific form of the native structure~\cite{afinsen75}.

This view has been challenged by ideas from polymer physics~\cite{dill99,yee94}.
In particular, polymers with hydrophobic monomers when placed in a
polar solvent like water will  collapse to a configuration where the hydrophobic residues
are protected from the solvent in the core of the
collapsed structure. Similarly, protein folding can be viewed as  polymer collapse
driven by hydrophobicity. The question then arises, how much
of the observed secondary structure is a result of this non-specific collapse
process?

To examine the role of hydrophobic interactions in folding, coarse-grained models of
proteins have been developed, which reduce the 20 possible amino acid
monomers to two types: hydrophobic (H) and polar (P). Further simplification
is affected by using random walks on two or three dimensional lattices to represent chain
conformations.   Vertices of the lattice
visited by the walk are identified with monomers, which in the HP model are of the H or P variety.
Furthermore, in order to capture the compact nature of the folded protein state,
Hamiltonian walks are often used for chain conformations.
The Hamiltonian walk (or ``compact polymer'') is a self-avoiding walk on a lattice that
visits {\em all} the lattice sites.  The compact polymer model was first used by Flory
\cite{flory} in studies of polymer melting, and was later introduced by Dill \cite{dill99} in the
context of protein folding. The HP model provides a simple model within which a variety of
questions regarding the relation of the space of sequences (ordered lists of H and P monomers) to the space of
protein conformations (Hamiltonian walks) can be addressed; for a recent example see Ref.~\cite{li96}.

One of the first questions to be examined within the compact polymer model was
to what extent is the observed  secondary structure
of globular proteins (i.e., the appearance of well ordered helices and sheets) simply
the result of the compact nature of their native states. Complete enumerations
of compact polymers with lengths up to 36 monomers found a large average fraction of monomers participating in
secondary structure~\cite{chan89}. This added weight to the argument that the observed secondary structure
in proteins is simply a result of hydrophobic collapse to the compact state. This simple view was
later challenged by off-lattice simulations, which showed that specific local interactions among monomers are
necessary in order to produce protein-like helices and sheets~\cite{socci94}.

Here we reexamine the question of secondary structure in compact polymers on the square lattice
using Monte-Carlo sampling of the configuration space. We compute the probability of a
monomer participating in secondary structure in the limit of very long chains. We show that this
probability is strictly less than one, and that it depends on the precise definition of secondary
structure in the lattice model. We also show that, in the long-chain limit, compact
polymers are much more likely to exhibit secondary structure motifs than their non-compact
counterparts, such as ideal chains, described by random walks, or polymers in a good
solvent, modelled by self-avoiding random walks. The Monte-Carlo technique described below
can be easily extended to three-dimensional lattices and other models (such as
the HP model) that make use of Hamiltonian walks. In a forthcoming paper we further
demonstrate its utility in the context of the Flory model of polymer melting \cite{richunp}.

Hamiltonian walks on different lattices are  also interesting statistical mechanics models in their
own right, as their scaling properties give rise to new
universality classes of polymers. An unusual property of these walks is that
different lattices do not necessarily lead to
the same universality class. This lattice dependence is linked to
geometric frustration that results from
the constraint that Hamilton walks must visit all the sites of the lattice.  In
addition, compact polymers can be obtained as the zero-fugacity limit of fully packed loop
models (the exact form of which depends on the lattice) allowing for
the exact calculation of critical exponents~\cite{jacobsen98}.

Numerical investigations of  compact polymers are typically hampered by the need to
generate a sufficient number of statistically independent
compact configurations for the construction of a suitable ensemble.
It is not hard to see that
attempting to generate compact structures by constructing
self-avoiding random walks on a lattice would indeed be a problematic
endeavor; current state-of-the-art algorithms are essentially
``smarter'' chain growth strategies where the next step in the random
walk is taken based on not only the self-avoidance constraint but on a
sampling probability which improves as the program
proceeds~\cite{zhang03}.  Enumerations of all possible states have
been performed for both regular self-avoiding random
walks~\cite{liang02} and for compact polymers~\cite{camacho93}, but
this has only been possible for small lattices ($N < 36$).  Therefore,
an algorithm which can rapidly generate compact configurations on
significantly larger lattices, without the complication of
constructing advanced sampling probabilities would be an extremely
useful tool.

In Ref.~\cite{jernigan} a  method for generating compact polymers
based on the transfer matrix method was introduced. One limitation of the
method is that the transfer matrices become prohibitively large as the number
of sites in the direction perpendicular to the transfer direction increases above 10.
A very efficient Monte-Carlo method based on a graph theoretical approach was
introduced in  \cite{caruthers} and improved on in \cite{lua03} by reducing the
sampling bias.

One of the  purposes of this paper is to describe a Monte-Carlo
method  for efficiently generating compact polymer
configurations on the square lattice for chain lengths up to $N=2500$.
The Monte-Carlo algorithm outlined below makes use of  the ``back bite'' move, which was first
introduced by Mansfield  in studies of polymer melts \cite{Mansfield82}.
We perform a number of measurements to assess the validity and practicality of the algorithm for
generating compact polymer configurations. Probably the most important and certainly the
most elusive property is that of ergodicity, which would guarantee that the algorithm
can sample all compact polymer configurations. While we have been unsuccessful in constructing a proof of ergodicity,
we find excellent numerical evidence for it based on a number of different tests. In particular, we
check that the measured probability that the polymer endpoints are adjacent on the lattice is in agreement
with exact enumeration results for polymer chain lengths up to $N=196$. Furthermore, we demonstrate
that the  Monte-Carlo process satisfies detailed balance, which guarantees, at least in
the  theoretical limit of infinitely long runs, that the sampling is unbiased. We check this in practice
with a quantitative test of sampling bias for $N = 36$ (ie.~for compact polymers on a $6 \times 6$
square lattice).

For the Monte-Carlo process to be useful it should also sample the space of compact polymer configurations
efficiently. To quantify this property of the algorithm, we
measure the processing time required to generate a fixed number of compact polymer conformations, and find it to be
linear in chain length $N$.
Since the sampling is of the Monte-Carlo variety a certain number
of Monte-Carlo steps need to be performed before the initial and the final structure can be
deemed statistically independent. We find that this correlation time, measured in
Monte-Carlo steps per monomer, grows with chain length as $N^z$ with $z\approx 0.16$.

We put the Monte-Carlo algorithm to good use by tackling the question of the statistics of secondary structure in
compact polymer chains. While the previous study by Chan and Dill \cite{chan89} found a large
fraction of monomers participating in secondary-structure motifs, the polymer physics question
of what happens to this quantity in the long-chain limit, remained unanswered. Based on exact enumerations
for chain lengths up to $N=36$ the hypothesis that was put forward was that in the long chain limit
almost all the monomers will participate in secondary structure. Our computations on the other hand, show that the
probability of a monomer participating in secondary structure tends to a fixed number strictly less than one.
Furthermore the actual number depends on the precise definitions used for secondary structure
motifs. Still, from gathered statistics on the appearance of helix-like motifs in simple random walks
and self-avoiding walks,
we conclude that the propensity for secondary structures in compact polymers is much
greater than in their non-compact counterparts, even in the long chain limit. This provides further
support for the idea that the global constraint of compactness, imposed on globular proteins by hydrophobicity,
favors formation of secondary structure.

The paper is organized as follows. In section~\ref{MCsampling} we describe our Monte-Carlo process for
sampling compact polymers, which is based on Mansfield's backbite move \cite{Mansfield82}.
The correctness and usefulness of the Monte-Carlo algorithm for sampling compact polymer configurations,  is evaluated
in section~\ref{MCtest}. Finally, in section~\ref{SecStr} we give details of our computations of secondary
structure statistics for compact polymers with lengths ranging from $N=36$ to $N=2500$. The main conclusion of this
section is that the fraction of monomers participating in secondary structures is Gaussian distributed with a
variance that vanishes in the long-chain limit. Its mean is strictly less than one but still more than twice as
large as the values measured for non-compact lattice models of polymers.

\section{Monte-Carlo sampling of compact polymers}
\label{MCsampling}

\begin{figure}
\begin{center}
\includegraphics[scale = 0.40] {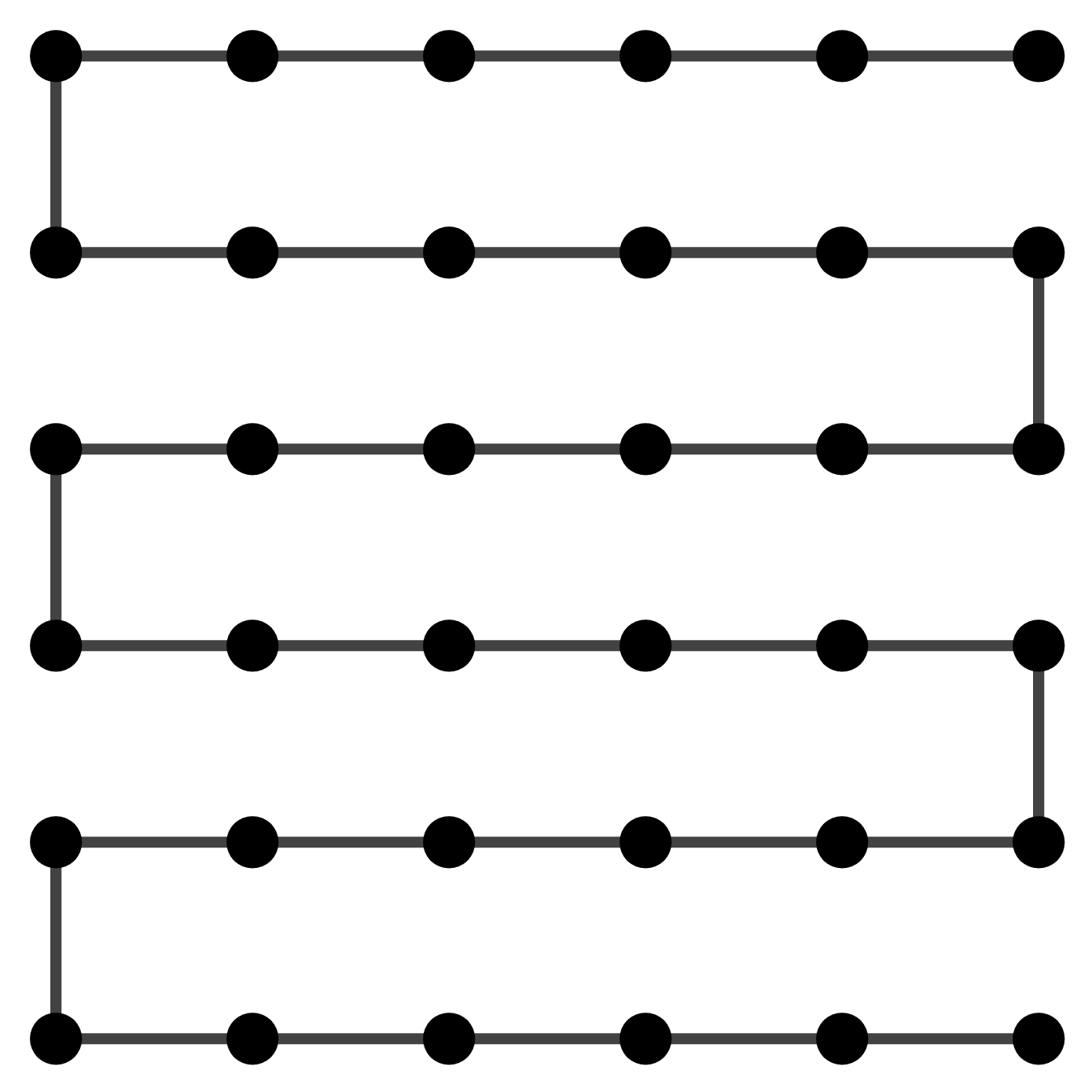}
\end{center}
\caption{Compact polymer configuration on a  $6 \times 6$ lattice. This ``plough'' configuration is
used as the initial state for the Monte-Carlo process.}
\label{init}
\end{figure}

\begin{figure}
\begin{center}
\includegraphics[scale = 0.35]{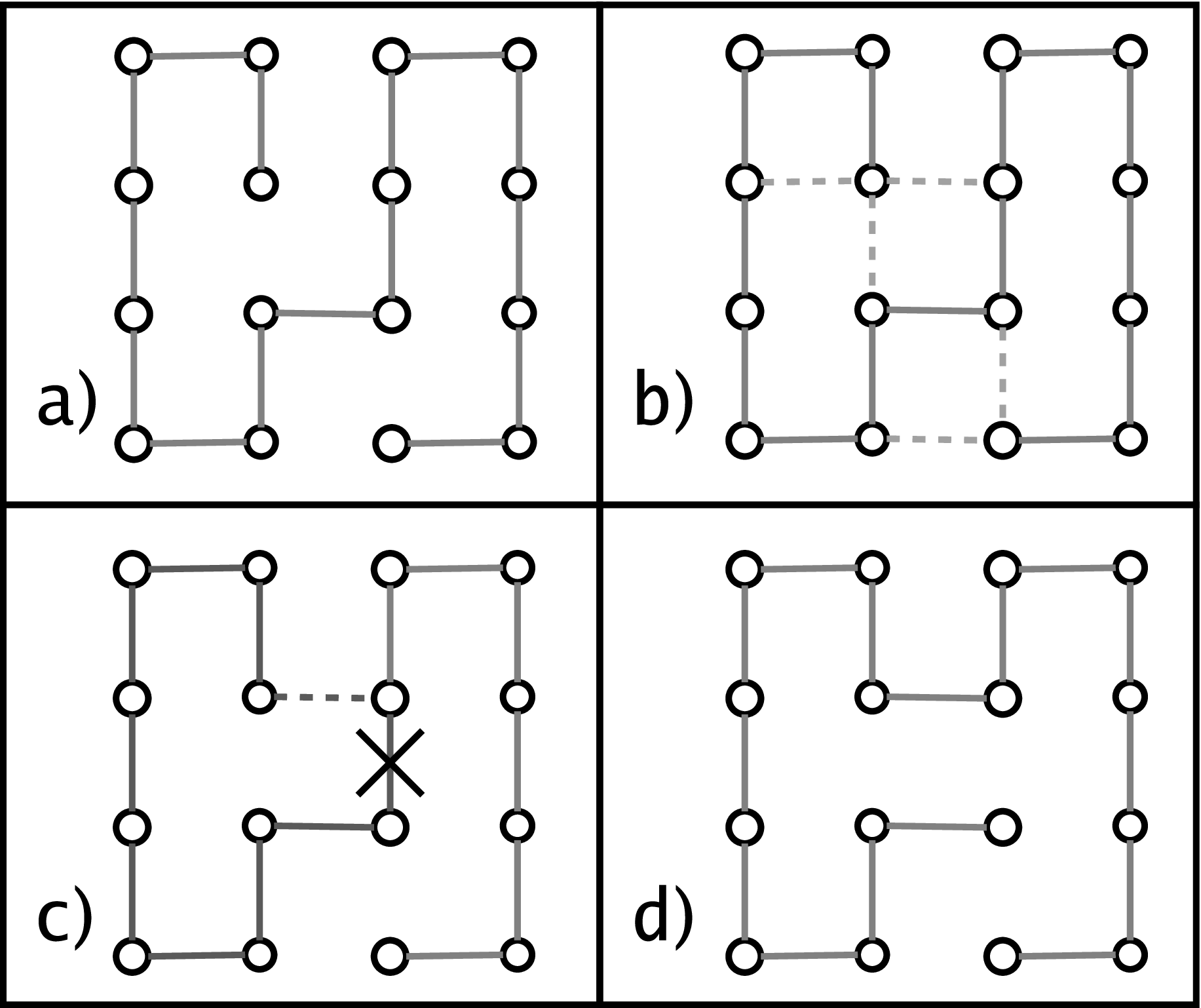}
\end{center}
\caption{Illustration of the ``backbite'' move used to generate a new Hamiltonian walk from an initial one.
Starting from a valid walk a), one additional step is made starting at either of the two ends of the walk, b). Next
we delete a step, shown in c), to produce a new valid walk, d).}
\label{process}
\end{figure}

The Monte Carlo process starts with an initial Hamiltonian walk on the lattice. We use a
square lattice with side $\sqrt{N}$, $N$ being the polymer length. The
initial walk is the ``plough'' shown in Fig.~\ref{init}.
Starting from this initial compact polymer configuration, new configurations are generated
by repeatedly applying the backbite move \cite{Mansfield82}. Namely,  given a Hamiltonian walk
(Fig.~\ref{process}a), a link is added between one of the walk's free
ends and one of the lattice sites adjacent, but not connected, to that
end. This adjacent site is chosen at random with each possible
site having an equal probability of being chosen
(Fig.~\ref{process}b). After the new link has been added we no longer
have a valid Hamiltonian walk, since {\em three} links are now
incident to the chosen site.  To correct this we remove
one of the three links, which is {\em uniquely} characterized by
being part of a cycle (closed path) and not being the link just
added (Fig.~\ref{process}c). After one iteration of this process one
of the ends of the walk has moved two lattice spacings, and a new
Hamiltonian walk has been constructed (Fig.~\ref{process}d).

\begin{figure}
\begin{center}
\includegraphics [scale = 0.35] {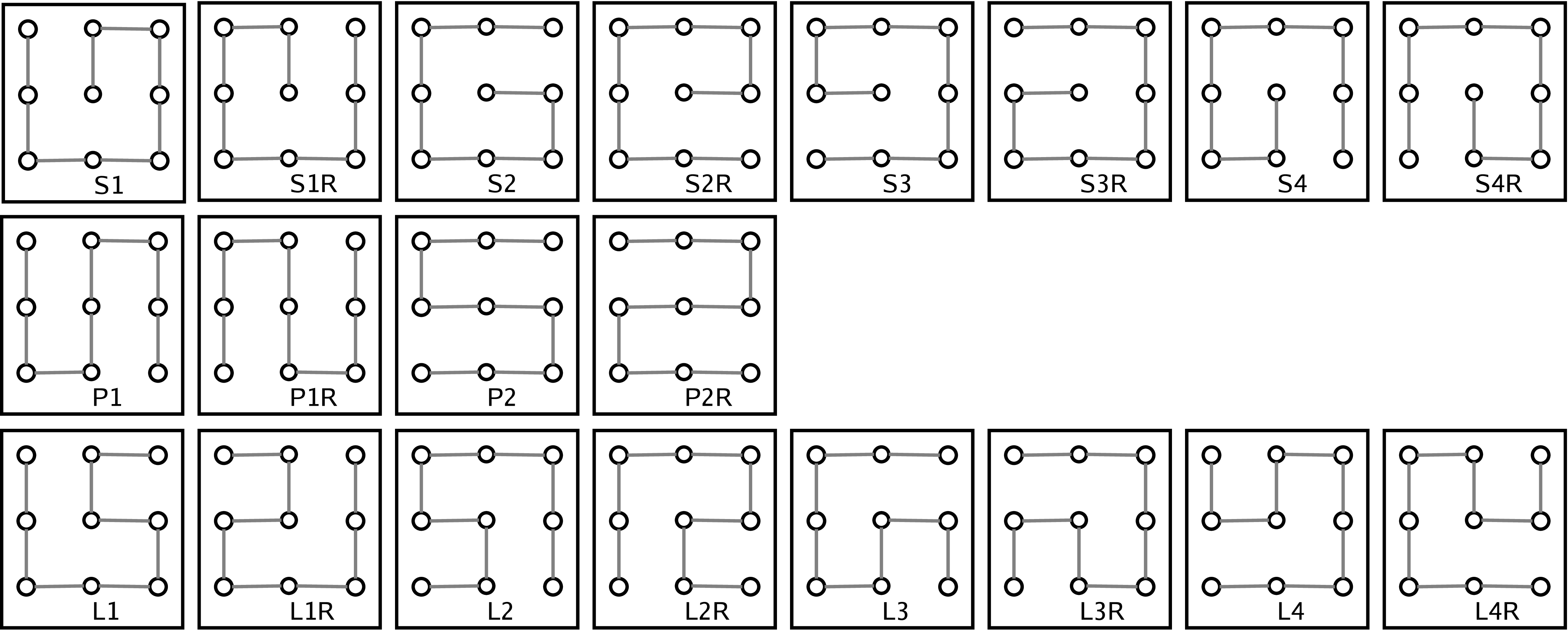}
\end{center}
\caption{Enumeration of all possible Hamiltonian walks on a $3 \times 3$ square lattice.}
\label{3by3enum}
\end{figure}

\begin{figure}
\begin{center}
\includegraphics [scale = 0.35] {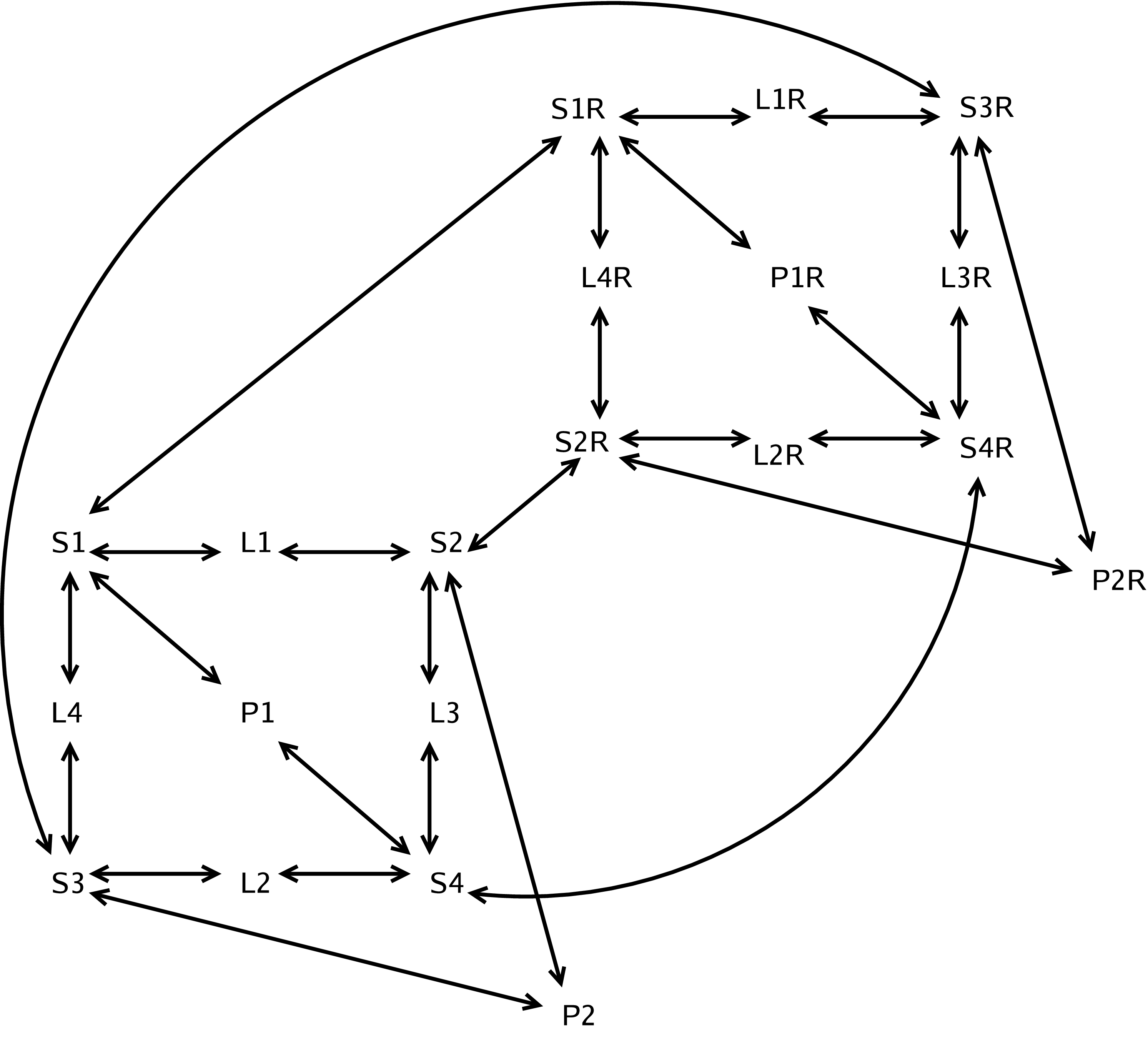}
\end{center}
\caption{Transition graph for compact polymers on the $3\times3$ square lattice generated
by the backbite move.}
\label{3by3trans}
\end{figure}

By repeatedly executing the backbite move it seems that
all possible Hamiltonian walks are generated. To examine this statement more closely,
we first consider compact polymers on a $3\times3$ lattice.
Fig.~\ref{3by3enum} shows an enumeration of all possible compact polymer configurations
on this lattice. The corresponding walks may be
divided into three classes where all the walks in a given class ({\bf
P}lough, {\bf S}piral or {\bf L}ocomotive---denoted P, S and L in the
figure) are related by reflection (denoted R in the figure) and/or
rotation. (Note that P-class walks are invariant under rotation by 180 degrees, and
that there are half as many P-class walks as S or L-class walks.)
 Fig.~\ref{3by3trans} shows the transition graph that connects compact polymer
 configurations on a $3\times 3$ lattice that are related by a single backbite move.
We see that all the 20 possible walks can be reached from any initial
walk. Furthermore, it is important to notice that the S-class walks have four
moves leading in and out of them, while the L and P-class walks
only have two moves leading in and out of them. This happens
as a result of the locations of a walk's end points. Namely, on a square
lattice an end point on a corner can only be linked to one adjacent
site by the backbite move, end points on the edges can be
linked to two sites, while end points in the interior  of the lattice can be
linked to three sites. Because there are twice as many
moves leading to S-class walks as there are for P-class or
L-class, the S-class walks are twice as likely to be generated if
backbite moves are repeatedly performed (this subtlety is absent if periodic
boundary conditions are employed).

In order to compensate for this source of bias in
sampling of compact polymers, an adjustment to the original process is made:
for structures which have fewer paths available to access them, we
introduce the option of leaving the current walk unchanged in the next
Monte-Carlo step. The probability of making a transformation from the
current walk is calculated by counting how many links $l$ can
be drawn from the end points of the current walk and dividing that number by
the maximum number ($l_{\rm max}$) of links that could be drawn for any walk
on the lattice. For example, consider a P-class walk on a $3\times3$
lattice. There are two possible links that could be drawn from the
end points of this walk, but there is a maximum of 4 links that could
be drawn (which happens in the case of S-class walks). Thus the
probability of making a backbite move is $\frac{l}{l_{\rm max}} =
\frac{2}{4} = 0.5$.

With this adjustment of the original Monte Carlo process, all walks accessible
from the initial walk will occur with equal probability, upon repeating the
algorithm a sufficiently large number of times. Technically speaking, the
amended algorithm satisfies {\em detailed balance}. In general, the criterion
for detailed balance reads $p_\alpha P(\alpha\to\alpha') =
p_{\alpha'}P(\alpha'\to\alpha)$, where $p_\alpha$ is the probability of the
system being in the state $\alpha$, and $P(\alpha\to\alpha')$ is the
transition probability of going from the state $\alpha$ to another state
$\alpha'$. In thermal equilibrium one must have $p_\alpha = Z^{-1} \exp(-\beta
E_\alpha)$, where $\beta$ is the inverse temperature, $E_\alpha$ is the energy
of the state $\alpha$, and $Z=\sum_\alpha \exp(-\beta E_\alpha)$ is the
partition function. In the problem at hand we have assigned the same energy
(say, $E_\alpha=0$) to all states, whence the criterion for detailed balance
reads simply $P(\alpha\to\alpha')=P(\alpha'\to\alpha)$.

Now suppose that the
state $\alpha$ can make transitions to $l_\alpha$ other states. (In the above
example, $l_\alpha=2$ for the P-class walks and $l_\alpha=4$ for the S-class
walks.) Then we can choose $P(\alpha\to\alpha')$ equal to
$\pi(\alpha\to\alpha')\equiv{\rm min}(1/l_\alpha,1/l_{\alpha'})$ for
$\alpha\neq\alpha'$. Define $\gamma(\alpha)=
\sum'_{\alpha'}\pi(\alpha\to\alpha')$, where the sum is over the $l_\alpha$
states $\alpha'$
which can be reached by a single move from the state $\alpha$. In order to make sure that
probabilities sum up to $1$, we must introduce the probability
$P(\alpha\to\alpha)=1-\gamma(\alpha)$ for doing nothing. Better yet,
we can eliminate the possibility of doing nothing by renormalizing the
Monte-Carlo time. Namely, let the transition out of the state $\alpha$
correspond to a Monte-Carlo time $1/\gamma(\alpha)$ and pick the
transition probabilities as $P(\alpha\to\alpha')=
\pi(\alpha\to\alpha')/\gamma(\alpha)$. Then the transition {\em rates}
(i.e., the transition probability per unit time) satisfies detailed
balance as it should.  This renormalized dynamics is clearly optimal
in the sense that now the probability of leaving the state unchanged
is zero, $P(\alpha\to\alpha)=0$. In practice, the optimal choice only
presents an advantage if the numbers $l_\alpha$ are easy to evaluate (which is
the case here) and if their values vary considerably with $\alpha$ (which is
{\em not} the case here).  Accordingly, we have used only the simpler
$l/l_{\rm max}$ prescription described in the preceding paragraph.

Even though we have satisfied detailed balance, a walk generated by the Monte
Carlo process does not immediately start occurring with a probability that is
independent of the initial walk. For large $N$ in particular, a walk
generated by the process will show a great deal of structural similarity to
the walk that it was created from because only two links of the walk get
changed in each iteration of the process. To work around this problem a large
number of walks must be generated to yield the final ensemble. Below we address
this important practical issue in great detail.

\subsection{Properties of the Monte Carlo process}
\label{MCtest}

In evaluating the suitability of the Monte-Carlo algorithm for statistical studies of compact polymers
the following issues must be addressed:
1.~Does the process generate all possible Hamiltonian walks on a given lattice?
2.~Is the sampling as described in the previous section truly unbiased?
3.~How rapidly do descendant structures lose memory of the initial structure?
4.~How does the processing time to generate a fixed number of walks scale with
the number of lattice sites?
The first two questions relate to issues of ergodicity and detailed balance which both
need to be satisfied so that structures are sampled correctly. The last two questions
pertain to the efficacy with which the algorithm can generate uncorrelated structures
that can be used in computations of ensemble averages.
Below we give detailed answers to these questions.

We have been unable to provide a general proof of
ergodicity, i.e., that the Monte-Carlo process can generate all possible Hamiltonian walks
on square lattices of arbitrary size.  However, we have observed that the process
successfully generates all of the possible walks on square lattices of
size $3\times3$, $4\times4$, $5\times5$, and $6\times6$. It should be noted that for $5\times5$ and
$6\times6$ lattices, all possible ``combinations'' of end point
locations are possible, while on smaller lattices
only walks with corner-corner, core-corner, and corner-edge
combinations are allowed. Whether endpoints are on edges, corners,
or in the bulk  of the lattice is important because it determines
how many links might be drawn from an endpoint which in turn
determines the probability of making a Monte-Carlo step away from the current
structure. Both $5\times5$ and a $6\times6$ lattices have $l_{\rm max}=6$,
which is the largest possible $l_{\rm max}$ on the  square lattice. In this sense, we
consider these two lattices representative of larger lattices.

It should also be noted that the algorithm is likely to exhibit parity
effects. This is linked to the fact that a square lattice can be
divided into two sublattices (even and odd). Namely, on a lattice of
$N$ sites, the two end points must necessarily reside on opposite
(resp.\ equal) sublattices if $N$ is even (resp.\ odd). To see this, note
that when moving along the walk from one end point to the other, the
site parity must change exactly $N-1$ times. In particular, only when $N$
is even can the two end points be adjacent on the lattice.
It is therefore reassuring to have tested ergodicity for
both $5 \times 5$ and $6 \times 6$ lattices.

\begin{figure}
\begin{center}
\includegraphics [scale = 0.45] {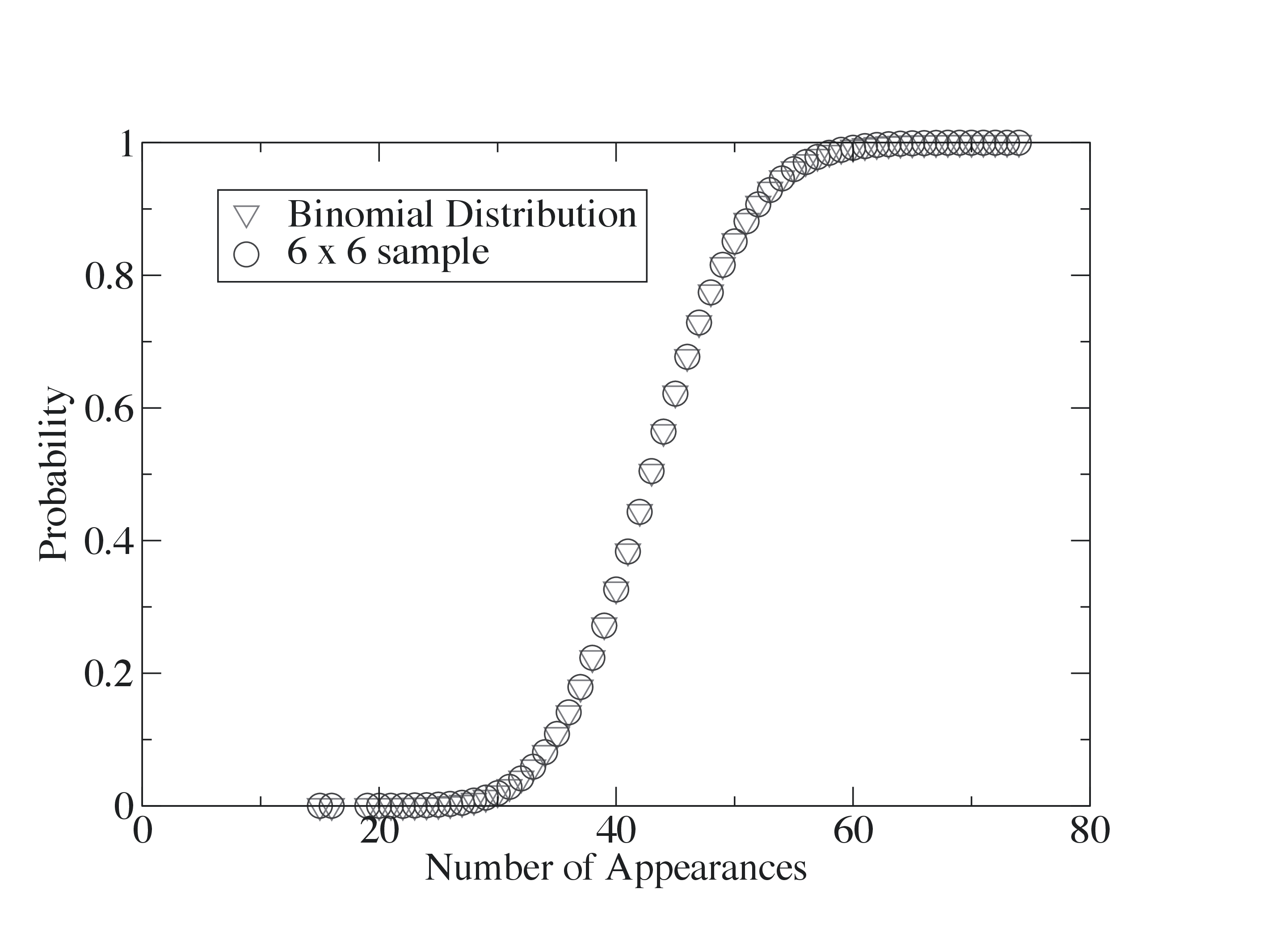}
\end{center}
\caption{Comparison of sampling statistics of compact polymers on the $6\times6$ lattice
produced by the Monte Carlo process with the binomial distribution, which is to be expected
for an unbiased sampling.}
\label{binomial}
\end{figure}

To test whether or not the process generates unbiased samples,
$K=10^7$ compact polymer conformations were generated on a $6\times6$ lattice. All different
conformations were identified and the number of occurrences of each identified conformation
was counted. The number of conformations on a $6\times6$ lattice is
known by exact enumeration to be $M = 229\,348$ \cite{unpub}, and
our algorithm indeed generates all of these.  Using a method similar
to the one used in Ref.~\cite{lua03} we construct the histogram of the
frequency with which each one of the $M$ possible conformations
occurs. This histogram is then compared to the relevant binomial
distribution. Namely, if each conformation occurs with equal
probability $p = \frac{1}{N}$ then the probability of a given
conformation occurring $k$ times in $K$ trials is $P(k) =
\frac{K!}{k!(K-k)!} p^{k}(1-p)^{K-k}$. In Fig.~\ref{binomial} we compare
$P(k)$ to the distribution constructed from the actual $6\times6$
sample. A close correspondence between the predicted distribution and
the distribution constructed from the Monte-Carlo data is evident
from the figure, indicating no detectable sampling bias in this
case.

\begin{table}
 \begin{tabular}{l|rrr}
   $L$ & $M_0/M_1$                   & $P_1^{\rm enum}$ & $P_1^{\rm MC}$ \\ \hline
  2   & $\frac{1}{4}$                & 1.00000000 & 1.0000 \\
  4   & $\frac{6}{276}$              & 0.34782609 & 0.3455\\
  6   & $\frac{1072}{229348}$        & 0.16826831 & 0.1664\\
  8   & $\frac{4638576}{3023313284}$ & 0.09819322 & 0.0979\\
  10  & $\frac{467260456608}{730044829512632}$
                                     & 0.06400435 & 0.0633\\
  12  & $\frac{1076226888605605706}{3452664855804347354220}$
                                     & 0.04488610 & 0.0442\\
  14  & $\frac{56126499620491437281263608}{331809088406733654427925292528}$
                                     & 0.03315399 & 0.0323\\
 \end{tabular}
 \caption{The probability $P_1$ that the walk's two end points are adjacent
 on an $L\times L$ lattice, as obtained by exact enumeration (see text) and
 by the Monte Carlo method.}
 \label{tabenum}
\end{table}

Further evidence that the sampling is unbiased is provided by computing
the probability $P_1$ that the end points of the generated walks are
separated by one lattice spacing. Note that when this is the case, the
walk could be turned into a closed walk, or Hamiltonian circuit, by adding
a link that joins the two end points. Conversely, a closed walk on an
$N$-site lattice can be turned into $N$ distinct open walks by removing
any one of its $N$ links. Therefore $P_1 = N M_0/M_1$, where $M_0$
(resp.\ $M_1$) is the number of closed (resp.\ open) walks that one can
draw on the lattice. Using this formula, we can compare $P_1$ as obtained
by the Monte Carlo method, to $P_1$ from exact enumeration data. The exact
enumerations are done using a transfer-matrix method for lattice sizes up to $14\times 14$
\cite{unpub}. The results
displayed in Table~\ref{tabenum} show that the two determinations of $P_1$
are in excellent agreement.

\begin{figure}
\begin{center}
\includegraphics [scale = 0.45] {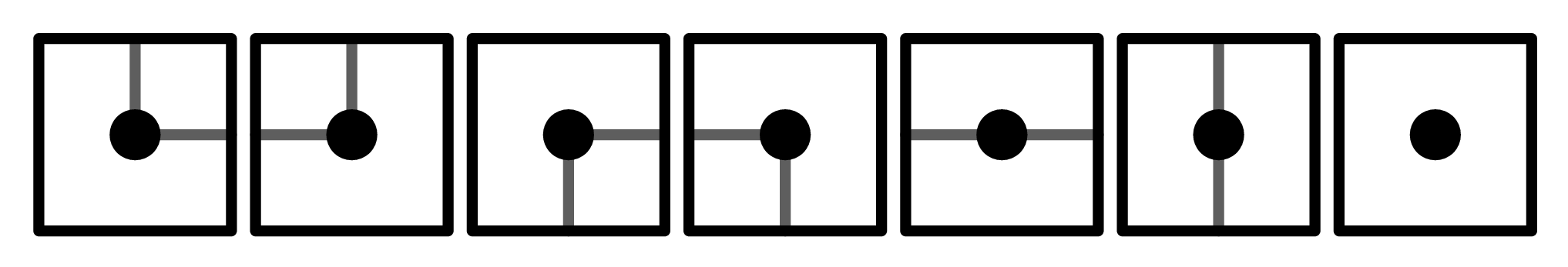}
\end{center}
\caption{Possible path shapes through the vertex of a square lattice.}
\label{paths}
\end{figure}

\begin{figure}
\begin{center}
\includegraphics [scale = 0.45] {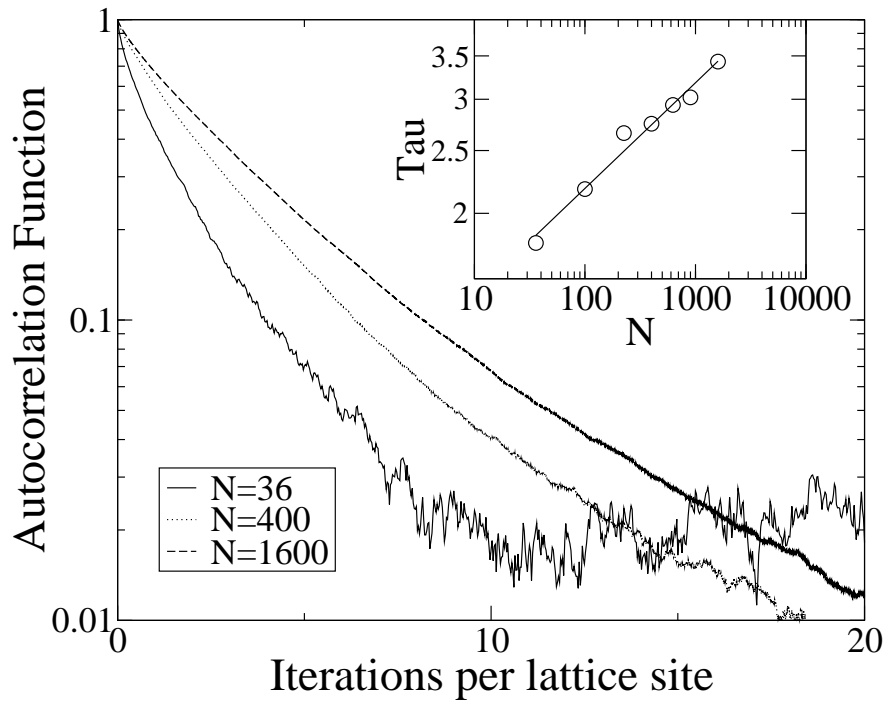}
\end{center}
\caption{Autocorrelation curves for various polymer lengths, $N$. This inset shows
the dependence of a characteristic time scale constant $\tau$, extracted from the autocorrelation
curves, on polymer length $N$.}
\label{decorrelation}
\end{figure}

To quantify the rate at which descendant structures become decorrelated from an initial
structure we must first devise a method for computing the similarity of two structures. The method
used here is to compare, vertex by vertex, the different ways in which the walk can pass through
a vertex. Looking at the examples of walks in Fig.~\ref{3by3enum} we immediately see that there are
seven possibilities, shown in Fig.~\ref{paths}. The walk may move vertically or horizontally
straight through a vertex, form a corner in four different ways, or terminate at a vertex. To
evaluate the autocorrelation time, a walk is represented by $N$ variables
$\sigma_i=1,2,\ldots,7$ that encode the possible shapes at each lattice site
$i$. The similarity $S(\sigma,\sigma')$ of two walks $\{\sigma_i\}$ and
$\{\sigma'_i\}$ is then defined as $S(\sigma,\sigma') = N^{-1} \sum_i
\delta_{\sigma_i,\sigma'_i}$. As the Monte-Carlo process progresses from an initial
Hamiltonian walk we expect that the similarity between the initial and the
descendent walks to decay with Monte-Carlo time. In order to estimate the
characteristic time scale $\tau$ for this decay, we plot in Fig.~\ref{decorrelation}
the autocorrelation function
\begin{equation}
\label{eq:autocorr}
A(t) = \frac{\left < S(\sigma(0), \sigma(t)) \right > - \left< S\right>_{\rm min}}{ 1 - \left< S\right>_{\rm min}}
\end{equation}
where the average is taken over many runs and $\left< S \right>_{\rm min}$ is the smallest average similarity computed for the
duration of the Monte-Carlo process, for a given $N$. The walk $\sigma(t)$ is one obtained after $t$ Monte-Carlo steps per lattice
site applied to the initial walk $\sigma(0)$.

The curves in Fig.~\ref{decorrelation} have an initial,
exponentially decaying regime. In this regime we fit them to the function $A \exp(-t/\tau)$
to obtain as estimate for the autocorrelation time $\tau$.
The inset of Fig.~\ref{decorrelation} shows the dependence of $\tau$ on polymer length $N$,
plotted on a log-log plot.
 Fitting now the polymer length dependence
using $\tau = B N^z$, we get the
estimate $z=0.16 \pm 0.03$ for the dynamical exponent. The fact that the
dynamical exponent is small tells us  that increasing the polymer length
in the simulation will not lead to a large increase in computational cost.

\begin{figure}
\begin{center}
\includegraphics [scale = 0.45] {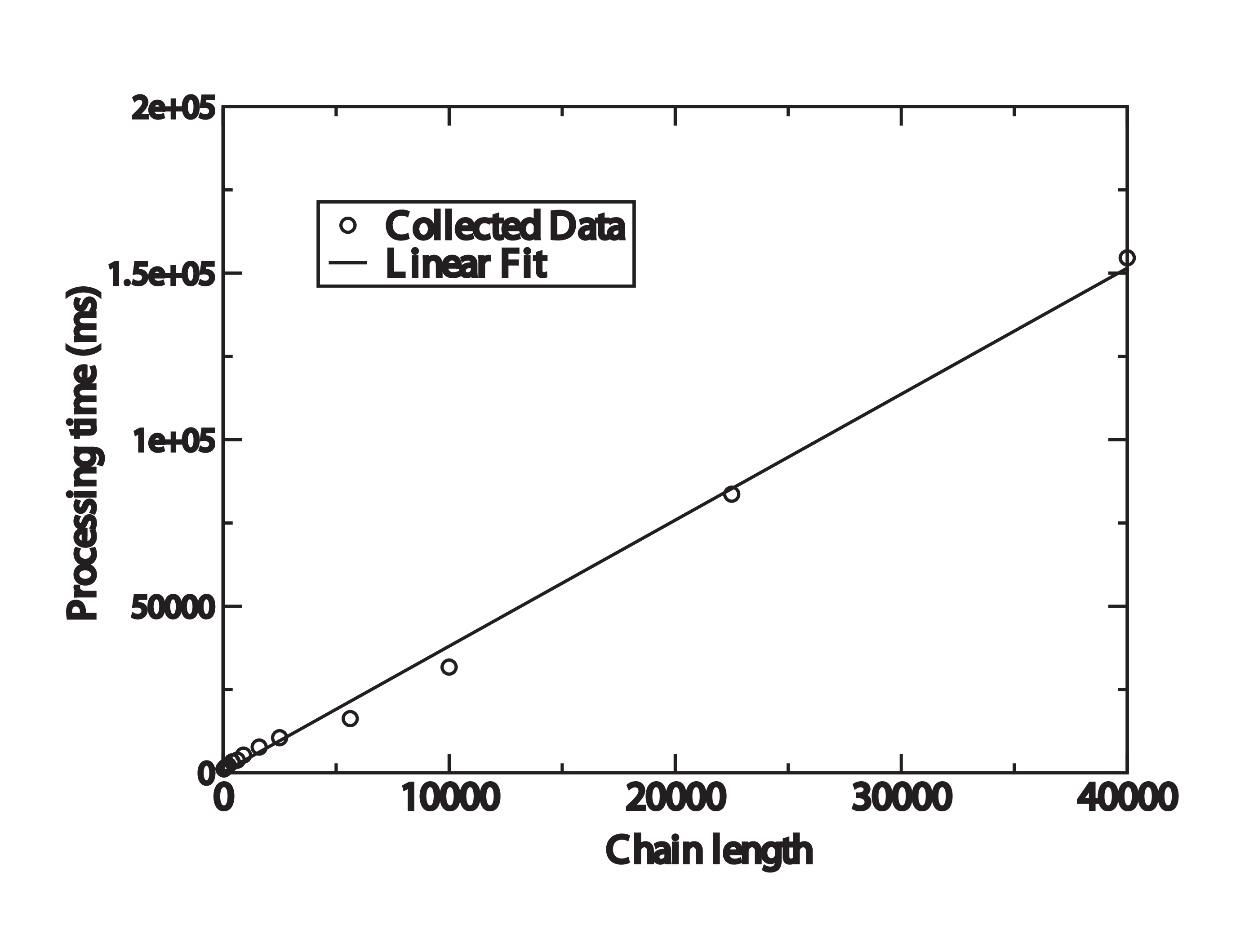}
\end{center}
\caption[]{Effect of varying the compact polymer length on processing time required
to generate $10^{6}$ consecutive structures by the Monte-Carlo process. Computations were performed on a
500 Mhz Pentium III computer with 512 megabytes of RAM.}
\label{ptime}
\end{figure}

There are two ways in which polymer length plays a role in the
performance of the algorithm described above. First, measurements show that the processor time needed
to generate a fixed number of walks scales linearly with their length
(see Fig.~\ref{ptime}). However, this particular result only
considers the time to generate a fixed number of consecutive
structures in the Monte-Carlo process, which, as we have seen, are not
statistically independent.
The actual processing time to generate an ensemble of properly sampled structures
would increase the reported times by a factor equal to  the number of iterations needed
to achieve statistical independence of samples. This factor roughly
equals the autocorrelation time $\tau$, which depends
on $N$ through the exponent $z$ determined above.

\section{Secondary structures in compact polymers}
\label{SecStr}

The presence of secondary structure-like motifs in compact polymers on the square lattice has
been extensively studied for chain lengths up to $N = 36$~\cite{chan89}. It
was shown in Ref.~\cite{chan89} that it is very unlikely to find a compact
chain with less than 50\% of its residues participating in secondary
structures and that the fraction of residues in secondary structures
increases as the chain length increases. Based on studies of chains up to $N =
36$ it appeared that the fraction of participating residues would
asymptotically approach 100\% as $N$ increased. Using the Monte-Carlo
approach described above we have extended these calculations to $N=2500$
and find that the fraction of residues participating in secondary
structures, in the long-chain limit, tends to a number strictly less than one.
We also show that this number is definition dependent but is still substantially greater
for compact polymers than for non-compact chains.

\subsection{Identification of secondary structures}

\begin{figure}
\begin{center}
\includegraphics [scale = 0.6] {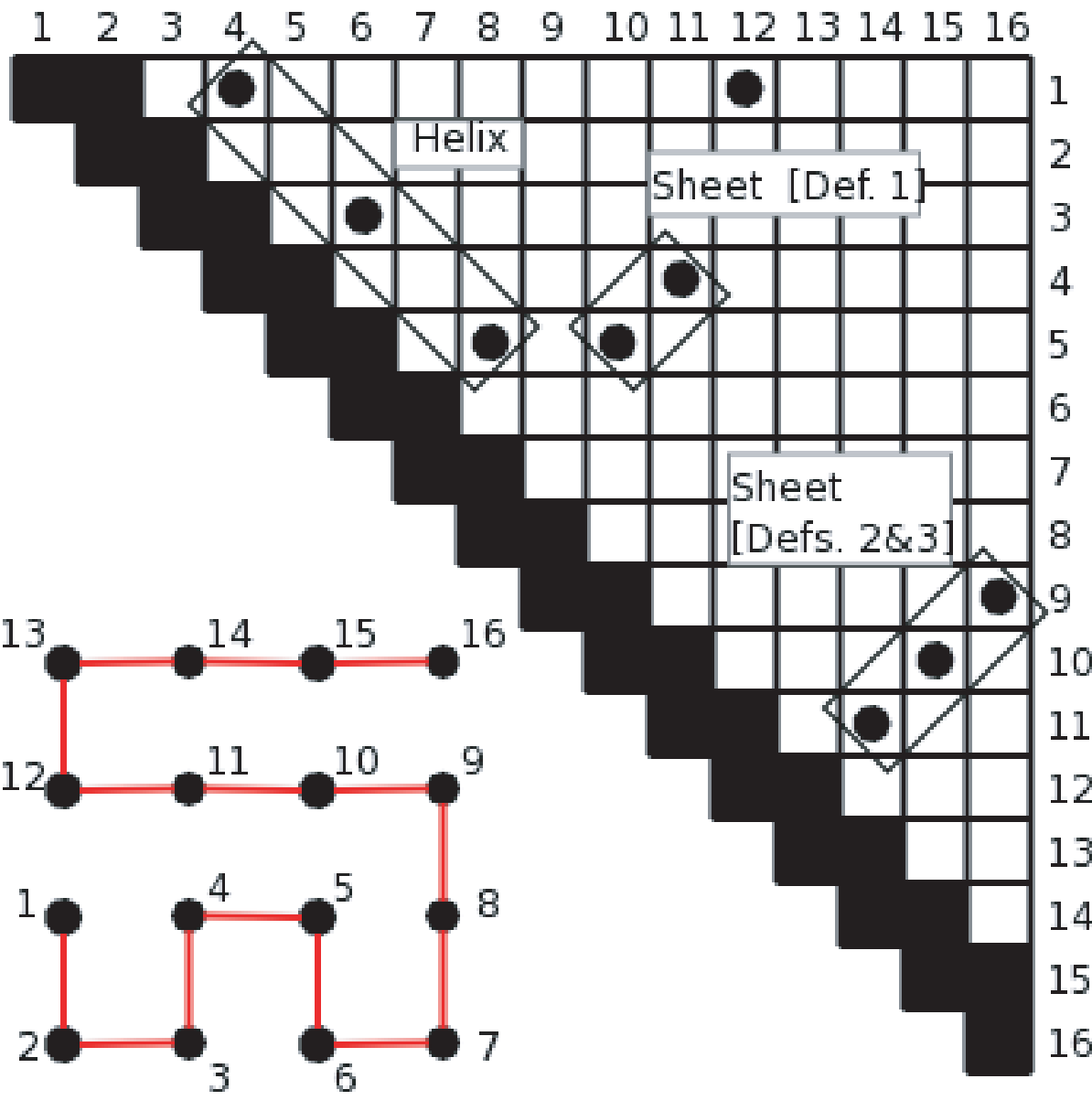}
\end{center}
\caption{Contact map for a Hamiltonian walk on a $4 \times 4$ square lattice. A filled circle in
position $(i,j)$ indicates that  residues $i$ and $j$ are in ``contact''; they are
adjacent on the lattice but are not nearest neighbors along the chain. Secondary structure
motifs defined in Fig.~\ref{ssdefs} appear as distinct patterns in the contact map. }
\label{cmap}
\end{figure}

There is more than one way to identify secondary structures in lattice models of
proteins. Following Ref.~\cite{chan89} we make use of contact maps which provide
a convenient and general way of representing secondary structure motifs.
A contact map is a matrix of ones and zeroes, where the ones represent
those pairs of
residues which are adjacent on the lattice, but not connected along the chain.
In this representation secondary structures are
identified by searching for patterns in the contact map which represent helices,
sheets, and turns; an example is shown in Fig.~\ref{cmap}.

\begin{figure}
\begin{center}
 \includegraphics [scale = 0.7] {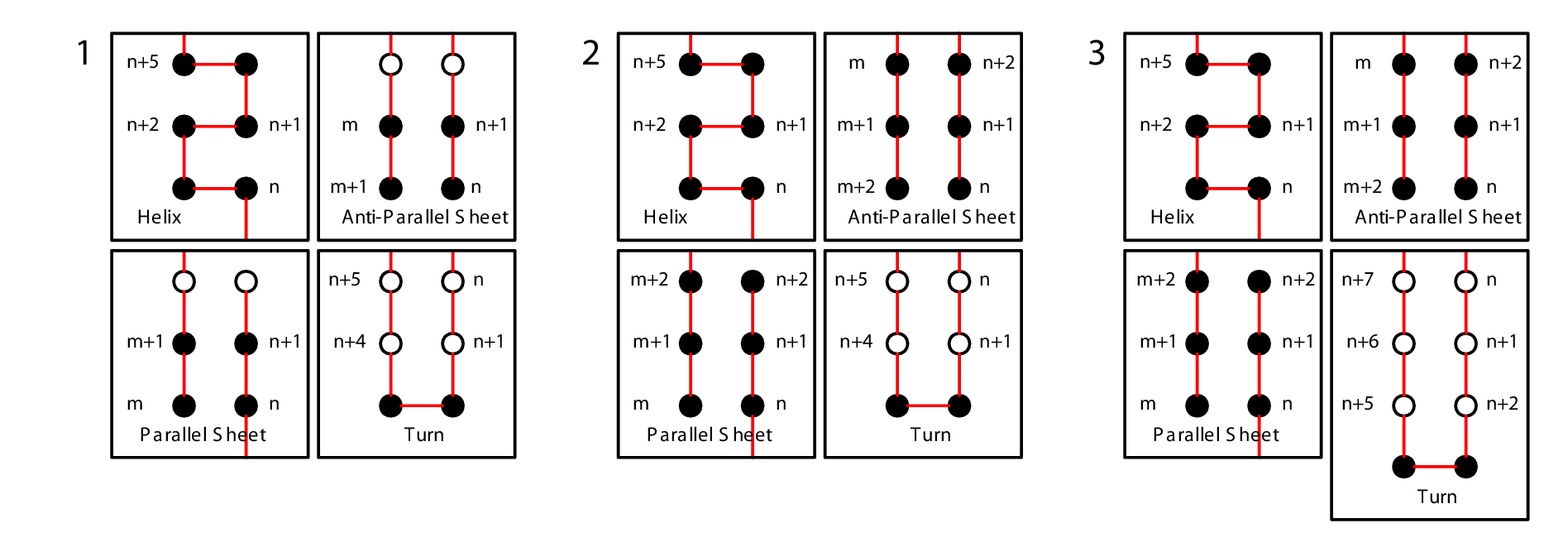}
\end{center}
\caption{Three definitions used to identify secondary structures in compact polymers. The shaded vertices (monomers)
are counted as participating in the particular secondary structure motif.   Definition
1 is the most liberal while 3 is the most conservative.
Definitions 1 and 2  are identical to those used in
Ref.~\cite{chan89}. The rationale for definition 3 is described in the text.}
\label{ssdefs}
\end{figure}

In order to test the generality of our findings, data was collected using
three different sets of definitions for secondary structure which are illustrated in
Fig.~\ref{ssdefs}. Since there is no unique definition of secondary structure for lattice
models of proteins, these models can at best provide qualitative answers to
questions relating to real proteins, like the role of hydrophobic collapse in
secondary structure formation. In other words, any conclusions derived from the
lattice model which might hope to apply to real proteins should certainly
not depend on the particular definition employed.

The first definition summarized in Fig.~\ref{ssdefs} is the least restrictive  one.
Because sheets only require two pairs of adjacent residues, this definition
allows for pairs of residues to participate in both helices and sheets. Unfortunately,
this property does not have any counterpart in real proteins. For this
reason, and following Ref.~\cite{chan89}, we also implement a second definition for
both parallel and anti-parallel sheets which requires them to have three pairs
of adjacent monomers instead of just two. This makes it more difficult for a
residue to be part of both a sheet and a helix. The third definition that we use, also shown
in Fig.~\ref{ssdefs},  is even more strict than the second definition: a turn now requires three pairs of
residues to be in contact. This ensures that a turn can only be
identified if it is part of a sheet, which was not necessarily the case in the second
definition.

\subsection{Statistics of secondary structures}

To gather the statistics on secondary structure motifs, 50\,000 statistically
independent Hamiltonian walks
on the square lattice were generated for chain lengths ranging between $N = 36$ and $N=2500$. For
each walk, the residues participating in secondary structure were identified
and counted using each of the three sets of definitions. To determine the
fraction of residues participating in secondary structure for a given walk,
the count is then divided by $N$, the total number of residues. The
histogram of the fraction of sites participating in secondary structure
is subsequently constructed for each chain length.

\begin{figure}[h]
\begin{center}
\subfigure[Definition 1, N=36]{\includegraphics[width = 0.45\textwidth]{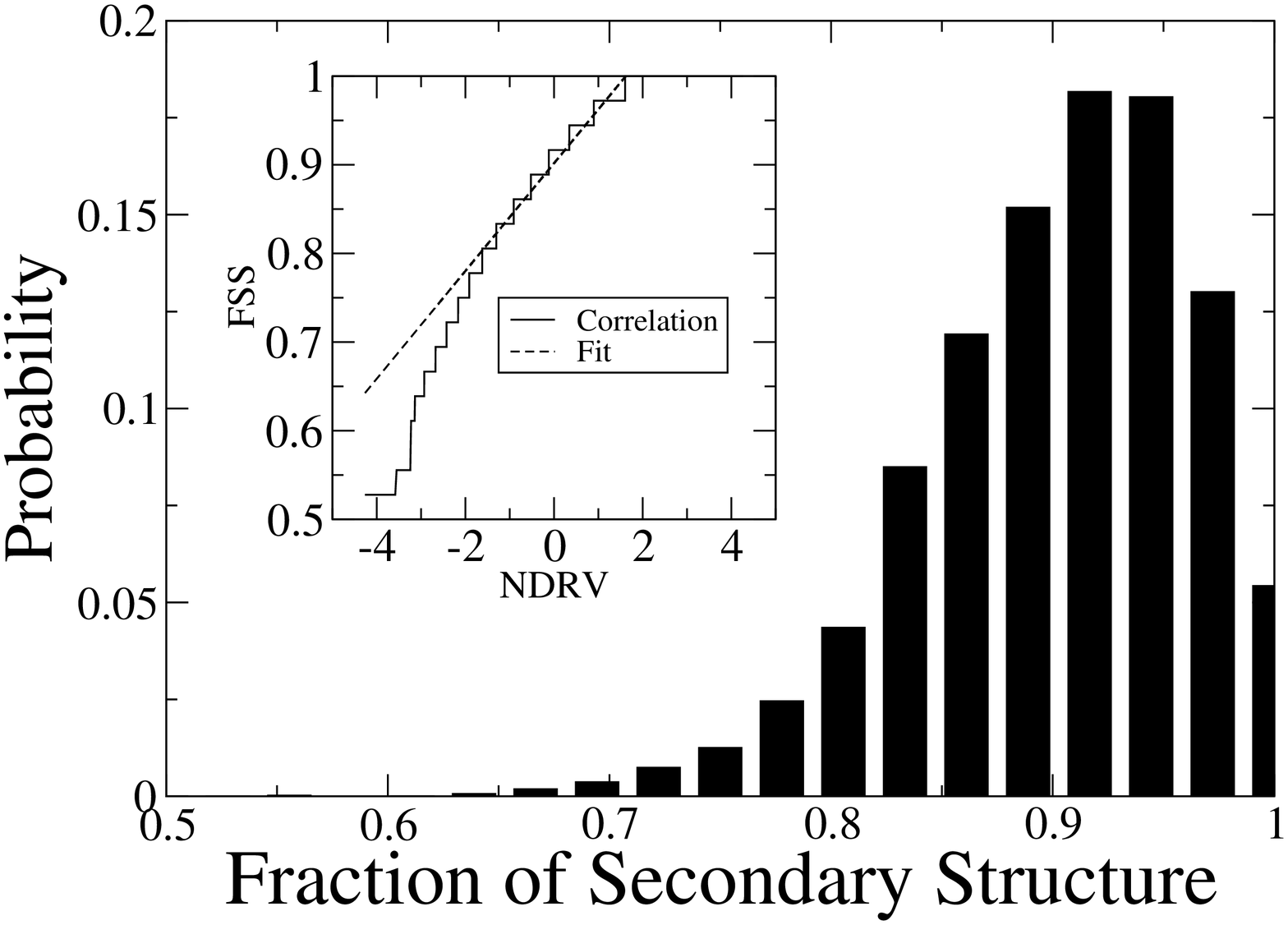}
\label{def1n36}}
\subfigure[Definition 1, N=400]{\includegraphics[width = 0.45\textwidth]{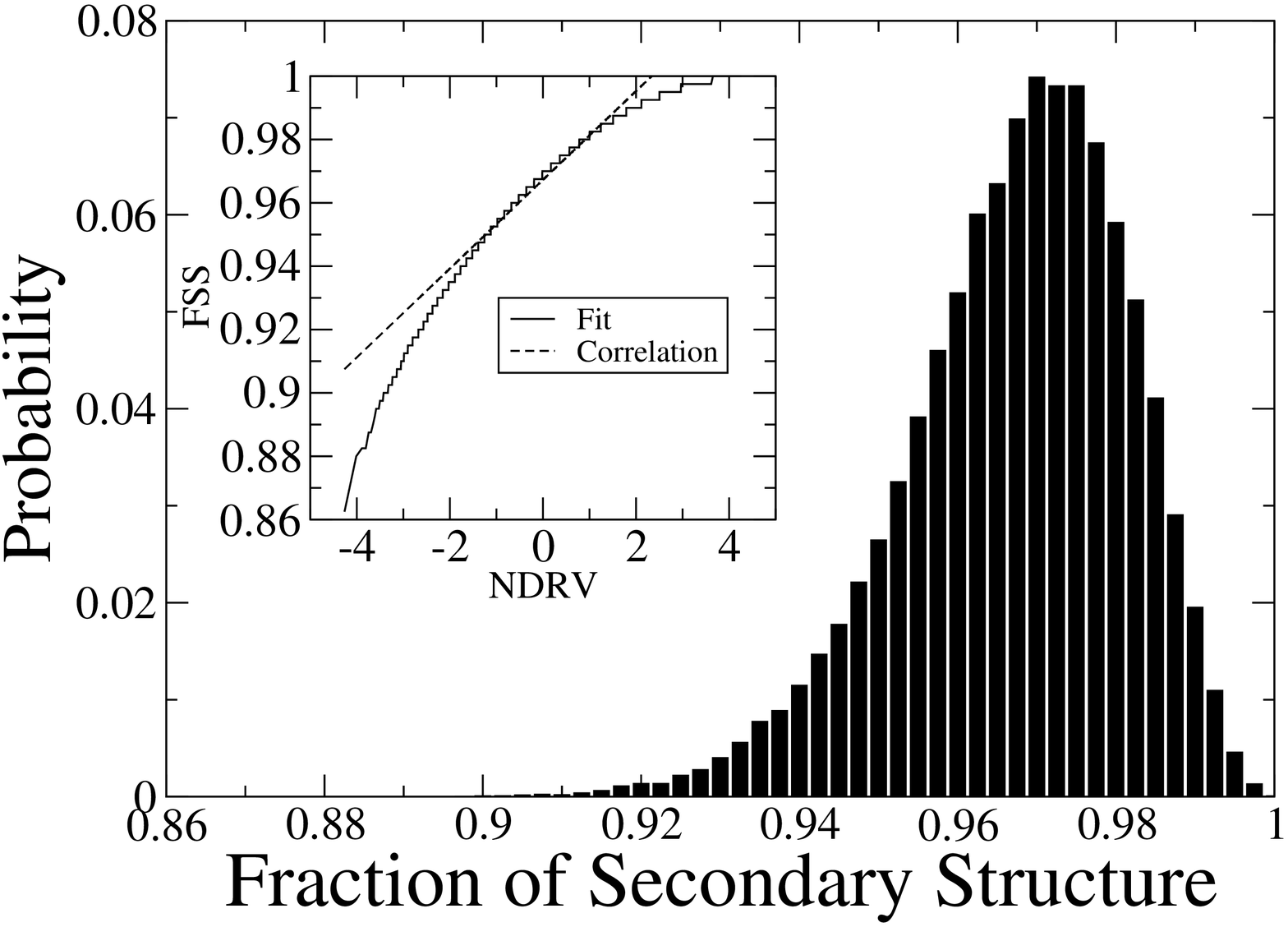}
\label{def1n400}}
\subfigure[Definition 1, N=2500]{\includegraphics[width = 0.45\textwidth]{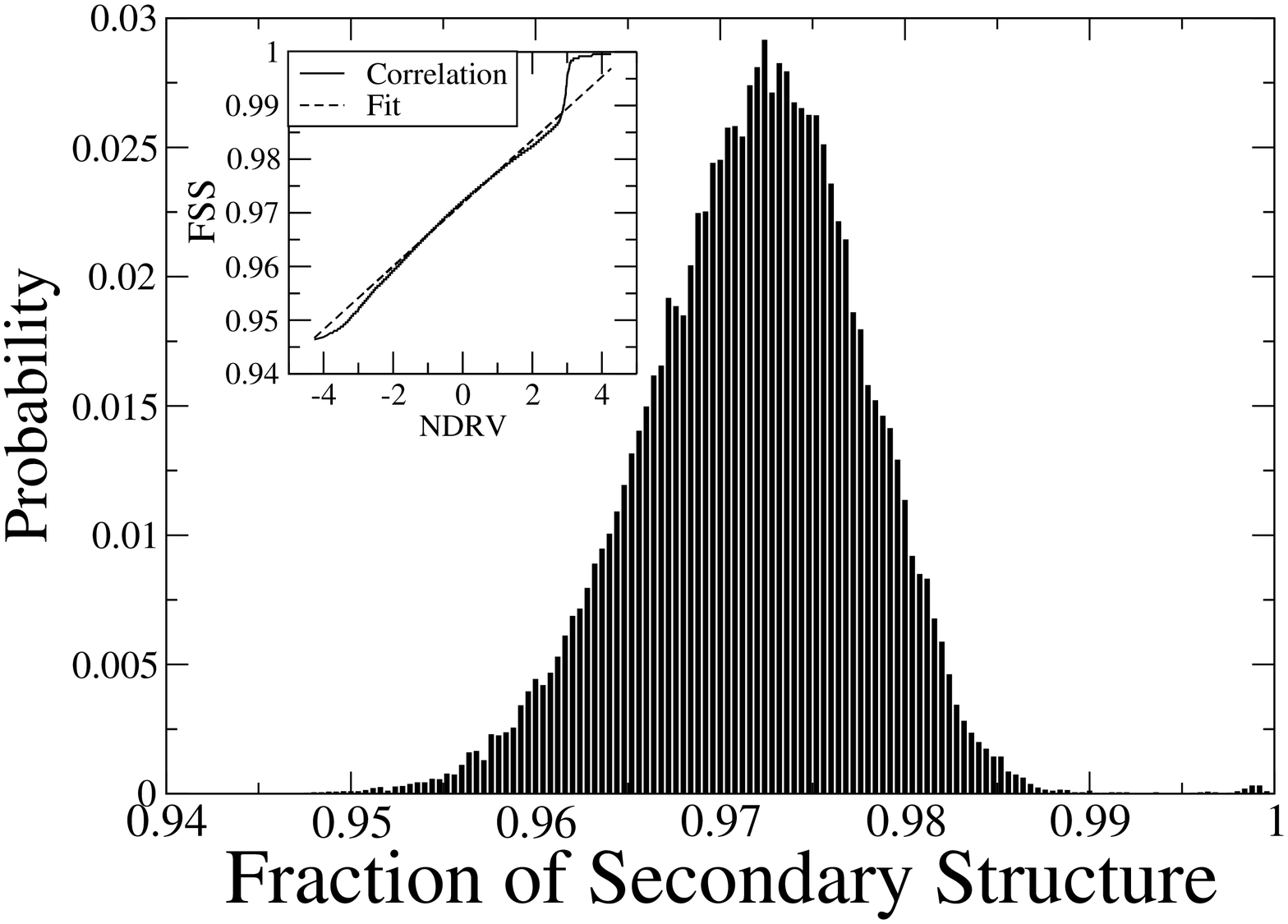}
\label{def1n2500}}
\subfigure[Definition 2, N=36]{\includegraphics[width = 0.45\textwidth]{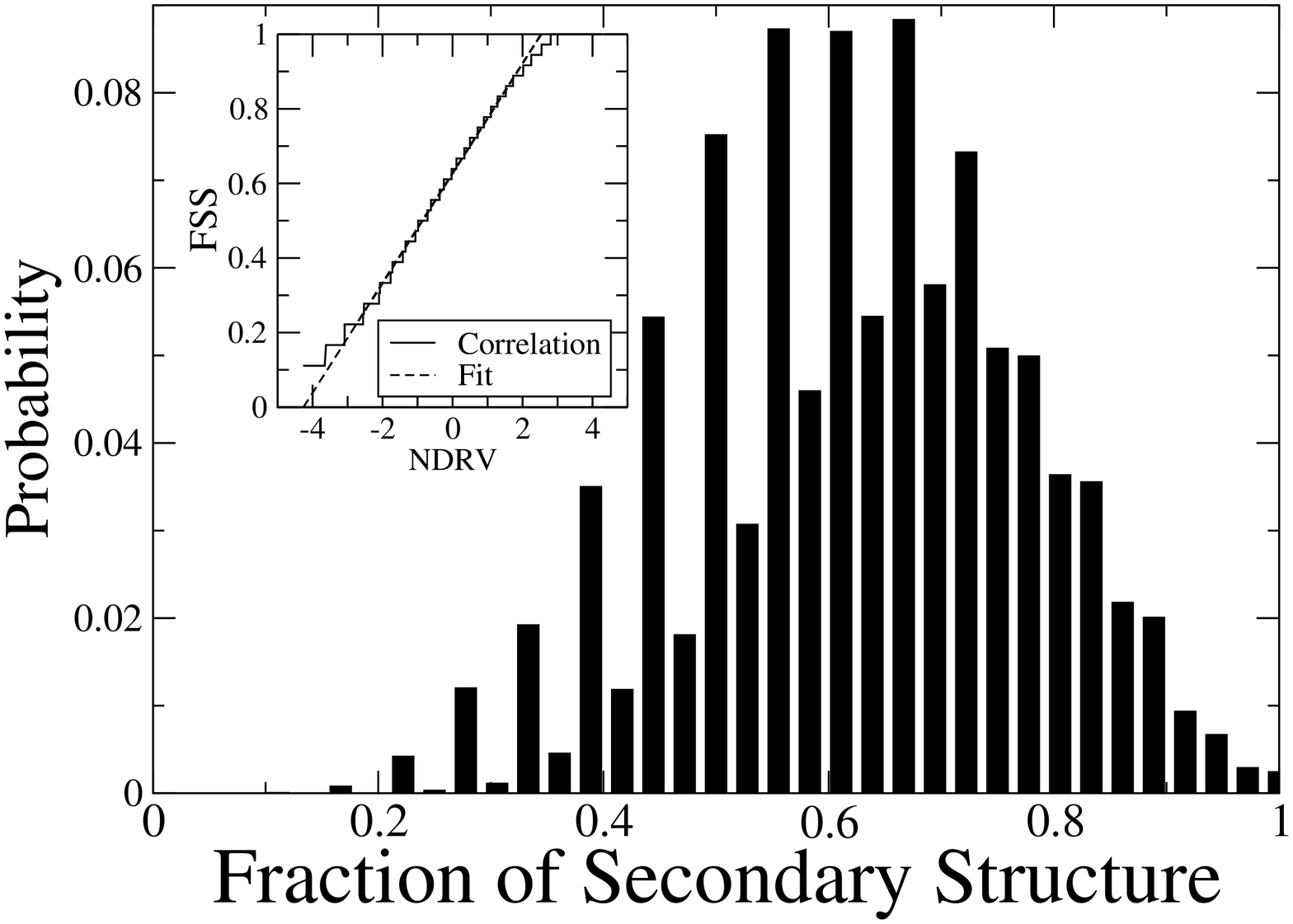}
\label{def2n36}}
\subfigure[Definition 2, N=400]{\includegraphics[width = 0.45\textwidth]{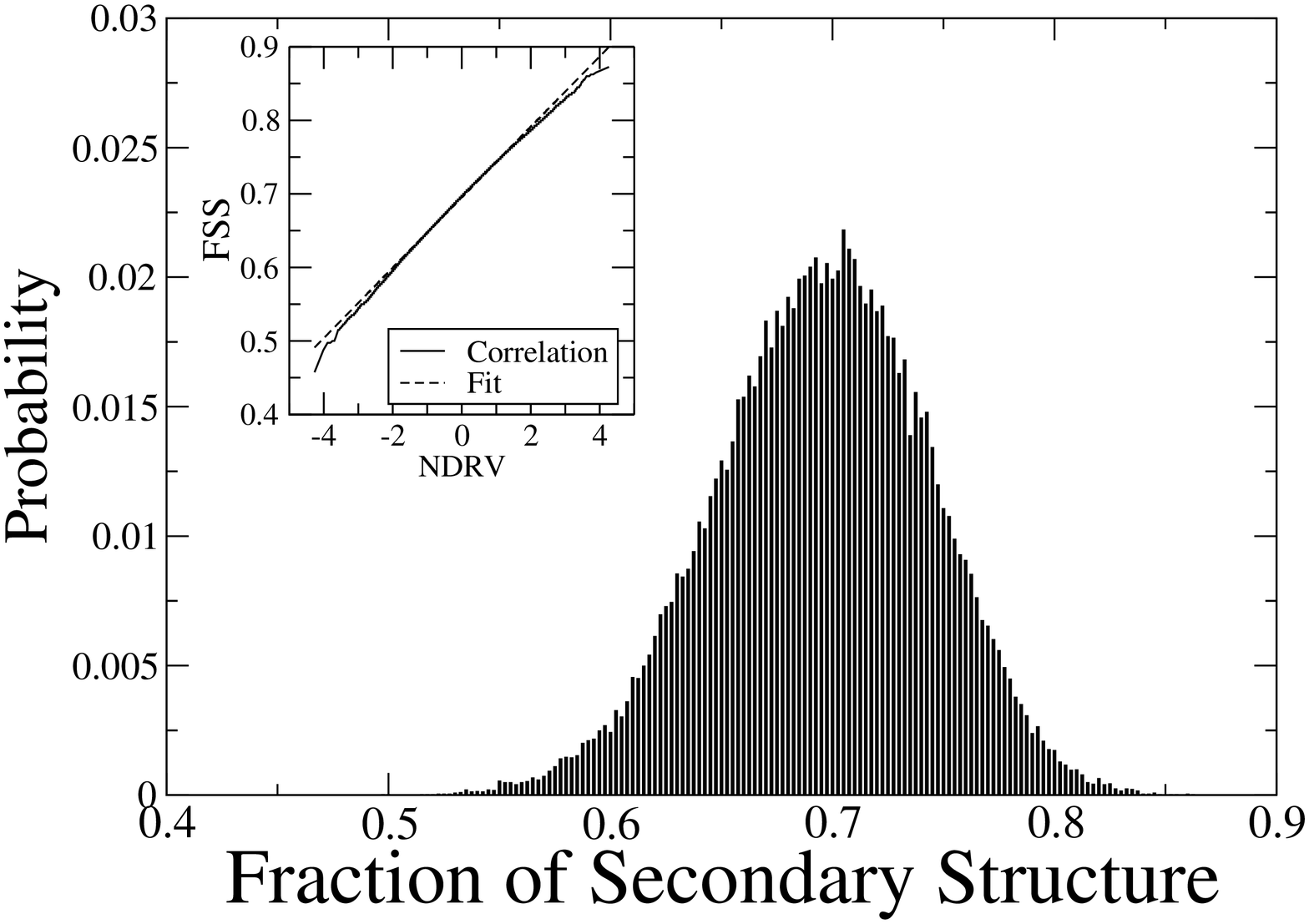}
\label{def2n400}}
\subfigure[Definition 2, N=2500]{\includegraphics[width = 0.45\textwidth]{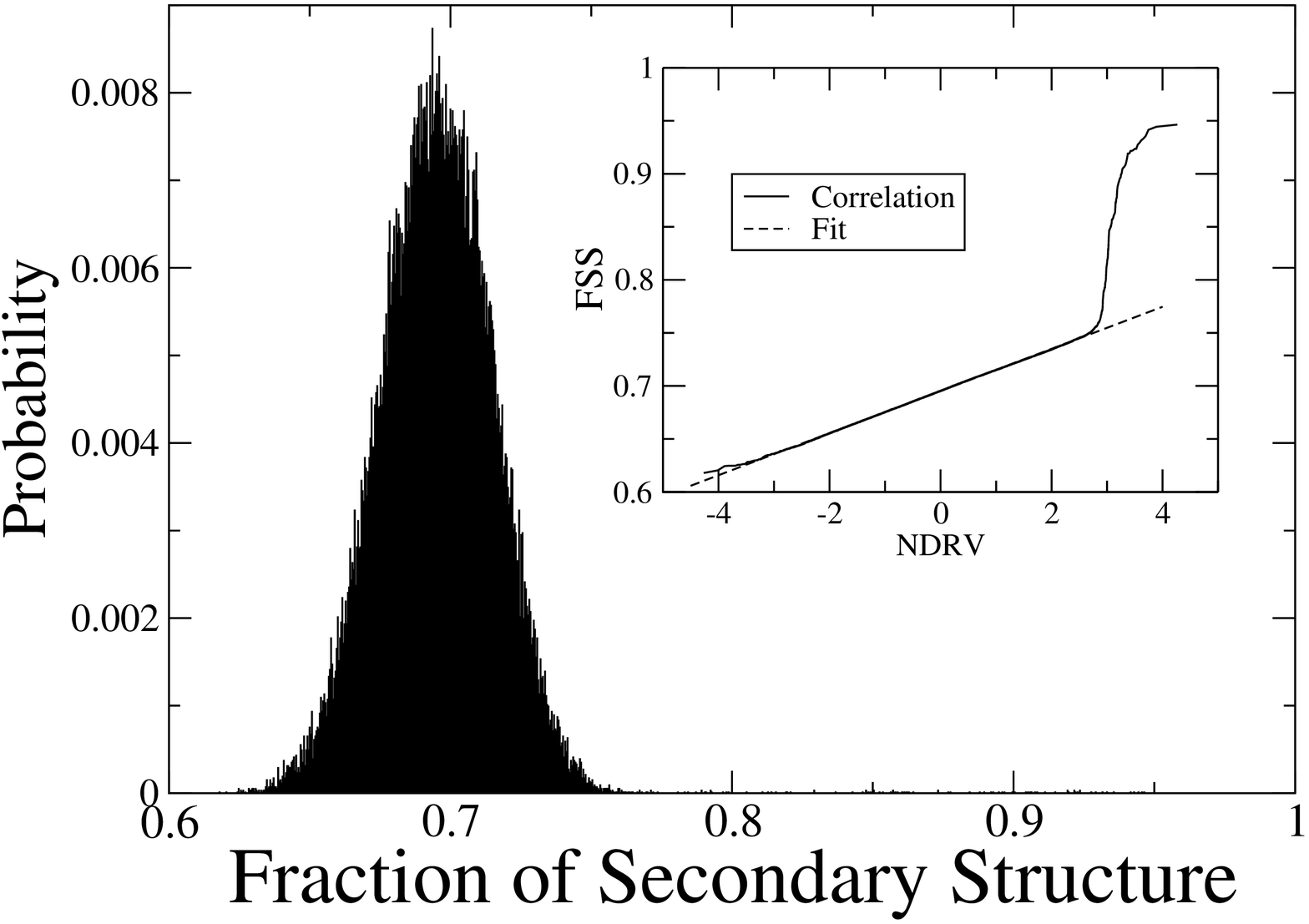}
\label{def2n2500}}
\end{center}
\caption{Probability distribution for the fraction of residues participating
in secondary structure for varying polymer lengths. The figures show only the first two definitions of
secondary structure represented in Fig.~\ref{ssdefs}. The insets
show the correlation between the measured distributions and a normally distributed random variable (NDRV). Straight lines
indicate a strong correlation to the Gaussian distribution and the slopes of the lines reflect the distribution
variances.}
\label{dist}
\end{figure}

Plots of the histograms of the participation fraction are shown in Fig.~\ref{dist} for
definitions 1 and 2. Both the mean and the variance of the participation fraction clearly depends on
the definition employed. As the polymer length $N$ increases the distributions appear to
approach a Gaussian shape for all definitions, and they are more and more sharply
peaked around the mean.

\begin{figure}
\begin{center}
\includegraphics [scale = 0.45] {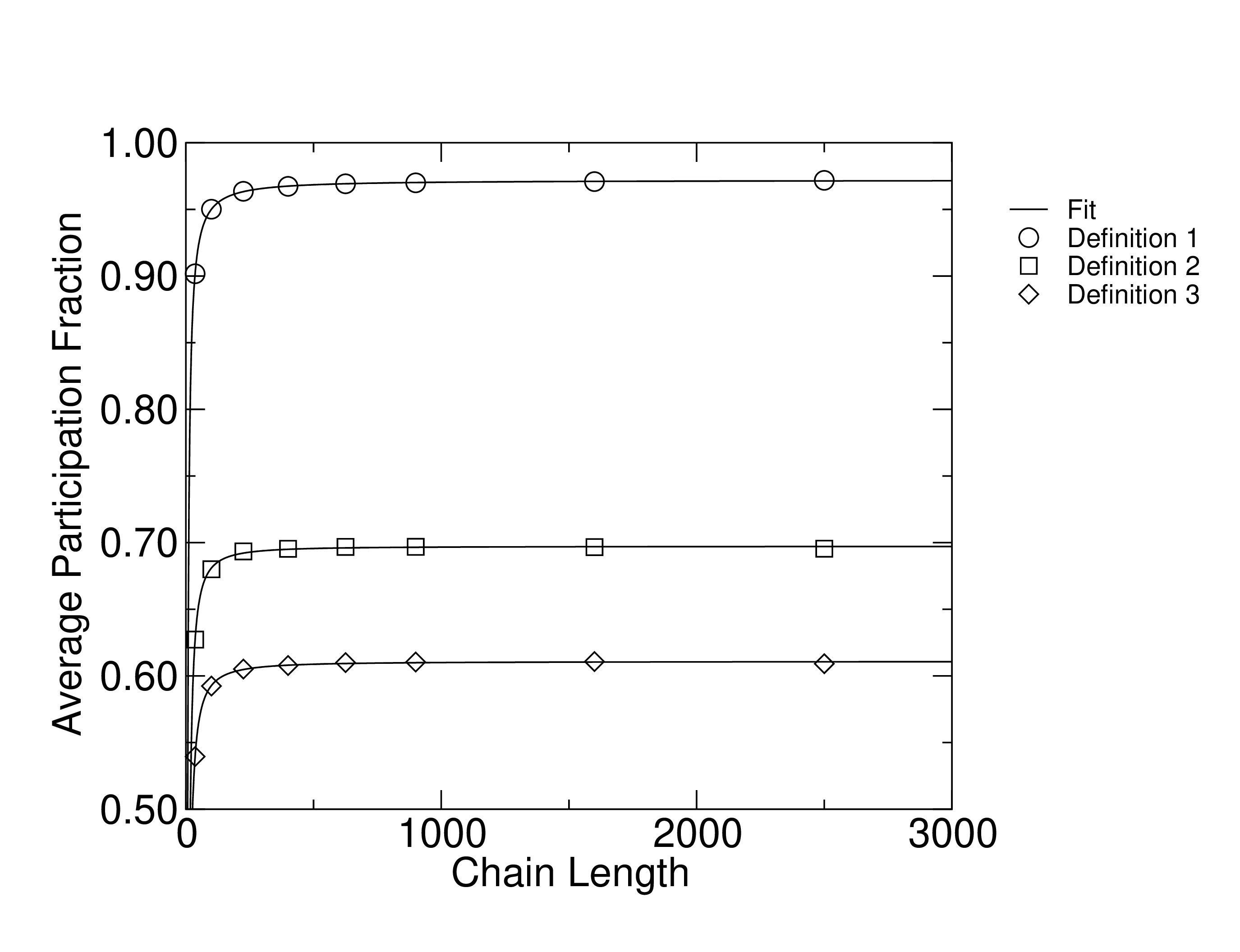}
\end{center}
\caption{Average fraction of residues participating in secondary structure as
a function of chain length $N$. The full lines represent a three-parameter fit to the function $f_\infty-a/N^x$.}
\label{percvsN}
\end{figure}

\begin{figure}
\begin{center}
\includegraphics [scale = 0.45] {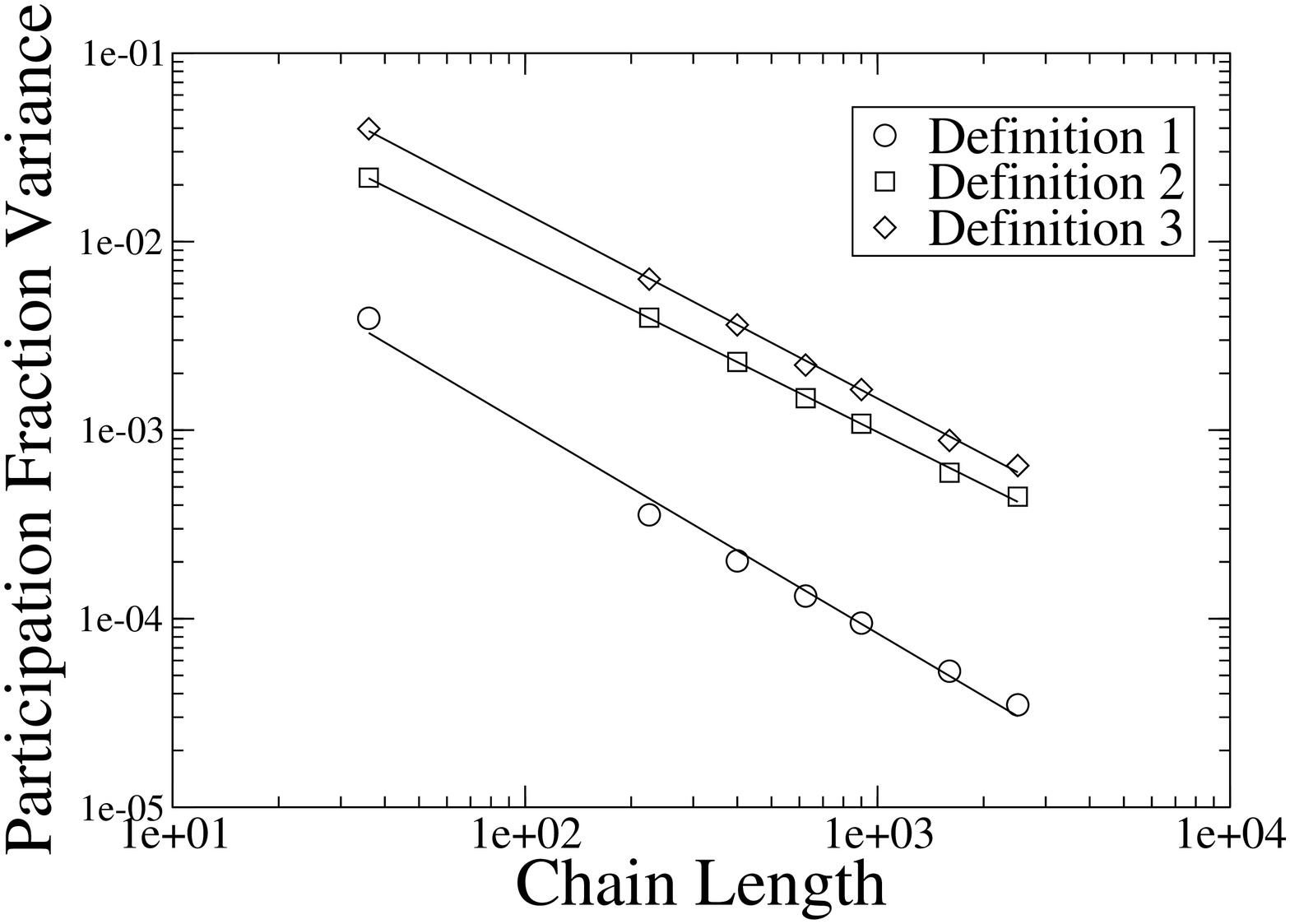}
\end{center}
\caption{Variance of the participation fraction  as a function of chain
length for all three definitions of secondary structure.}
\label{var}
\end{figure}

From the measured participation fractions we compute their mean and variance. The
dependence of the mean on the polymer length is shown in
Fig.~\ref{percvsN}. As polymer length increases the average fraction of
residues participating in secondary structure approaches a fixed number
$f_\infty$, which clearly depends on the definition used. Although the definition
affects the specific value of $f_\infty$, each curve has roughly the same shape
which is well fit by the function $f =f_\infty-a/N^x$. In all cases the numerical value of
$f_\infty$, the participation fraction in the long chain limit, is less than 1 (see Table~\ref{fitparms}).

\begin{table}
\begin{tabular}{l|rrr} Definition & $f_\infty$ & $a$ & $x$ \\
\hline
1 & 0.9719 & 4.2511  & 1.1451  \\
2 & 0.6972 & 11.7942 & 1.4303 \\
3 & 0.6108 & 8.8789  & 1.3461  \\
\end{tabular}
\caption{The parameters obtained from fitting the average fraction of residues participating in
secondary structures $f$ for different chain lengths $N$, to the functional form $f =f_\infty-a/N^x$.}
\label{fitparms}
\end{table}

The variance of the fraction of residues participating in secondary structure is shown
in Fig.~\ref{var}. It clearly decreases with $N$ in a power-law fashion. A linear
fit on the log-log plot reveals that the variance scales as $1/N$, regardless of the definition
of secondary structure employed. This result indicates that for compact polymers on the square
lattice the fraction of residues participating in secondary structure has a well defined
long chain limit given by $f_\infty$.

In order to quantify how closely the histograms in Fig.~\ref{dist} approach a  Gaussian distribution,
the percent of residues participating in secondary structure is plotted
against a normally distributed random variable. These plots appear as  insets in
Fig.~\ref{dist}, and a  straight line  indicates a Gaussian distribution.
Note that deviations from a straight line appear in the tails of
the distributions. We attribute this primarily to the influence of the initial
``plough'' configuration on the sampled walks. This we verified by comparing histograms for the participation
fraction constructed from three different ensembles of compact polymers
which differed by the number of Monte-Carlo steps taken before sampling is initiated.
As the initial wait time for the sampling to commence is increased we find that the
deviations from the Gaussian distribution decrease. In fact, in order to lose
memory of the initial plough state, we found the wait time to be of the order of $10 \tau$, where
$\tau$ is the measured correlation time.

In order to understand the degree to which global compactness, as opposed to local connectivity, of the
chains is responsible for the formation of helices we investigated the set of
all $2 \times 3$ motifs that can be observed in a compact polymer configuration.
Namely, on a $2 \times 3$ section of square lattice there are
7 possible bonds that can be drawn, which means there are $2^7$ different $2 \times 3$ motifs. Of course,
not all of these are compatible with a  compact polymer configuration. For example, motifs with all bonds
present or no bonds present could not be part of a valid Hamiltonian walk. In fact we found 67 allowed motifs, of which
only two are  helices.  Therefore, the naive assumption that each of the allowed motifs appears with an equal
probability would lead to the expectation of only 3\% of residues participating in helix motifs.
By comparison, simulations of long chains place the expected value
near 28\%.

\begin{figure}[h]
\begin{center}
\subfigure{a)\includegraphics[width=0.3\textwidth]{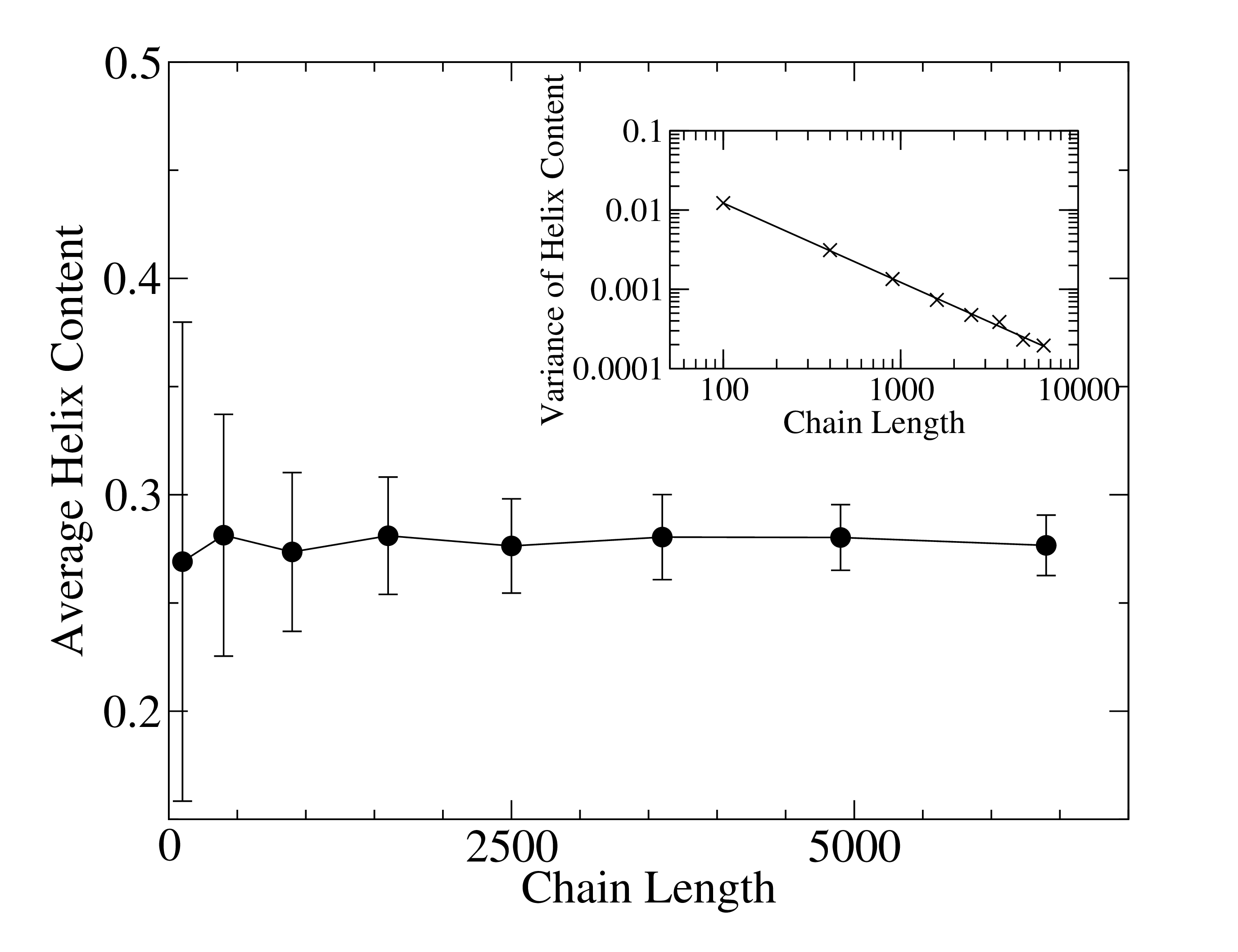}}
\subfigure{b)\includegraphics[width=0.3\textwidth]{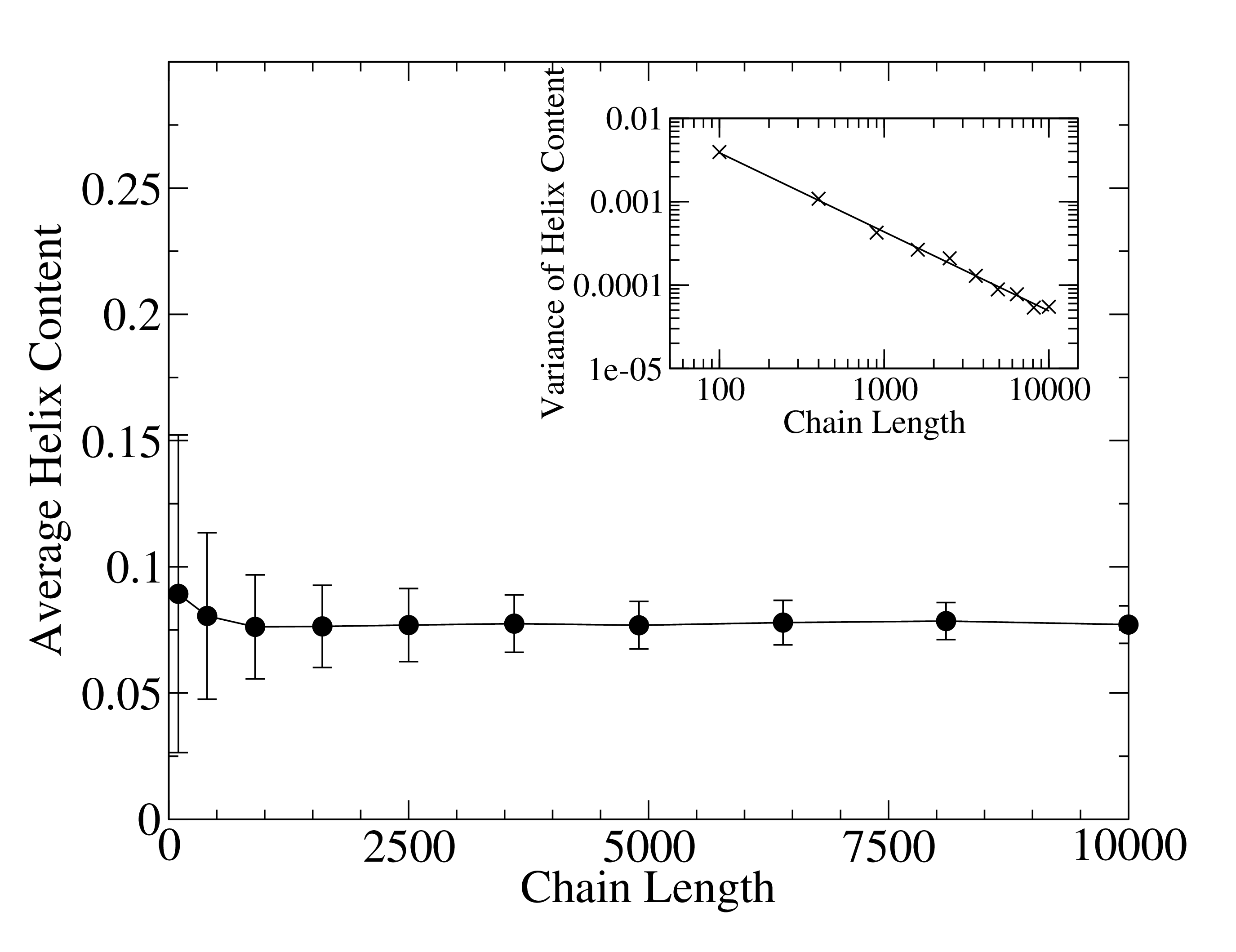}}
\subfigure{c)\includegraphics[width=0.3\textwidth]{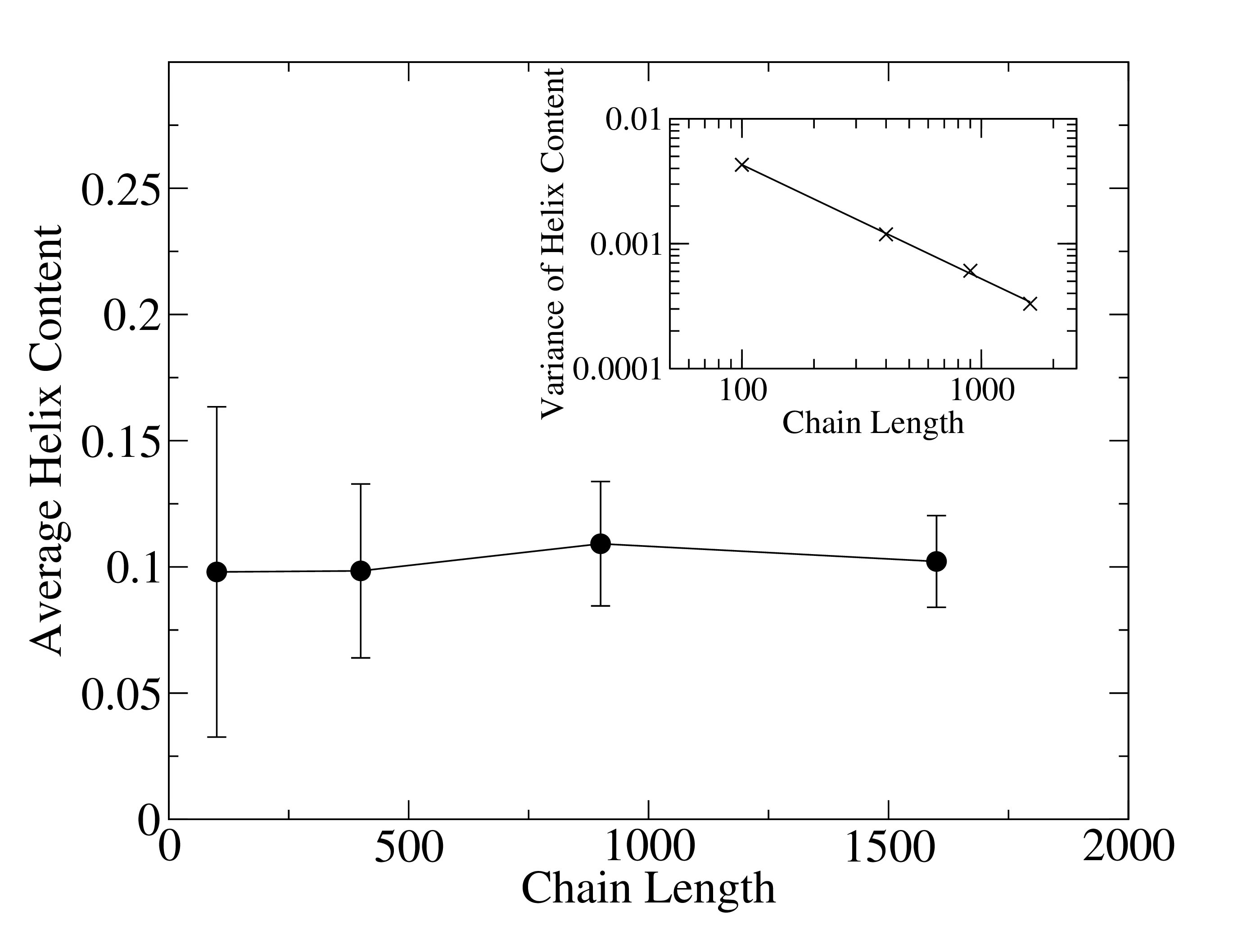}}
\end{center}
\caption{Average fraction of residues participating in helices for a) Hamiltonian walks,  b) random walks and
c) self-avoiding random walks. The insets show the variance of the helix content as a function of the
chain length. }
\label{cprwsaw}

\end{figure}

To further assess the importance of being compact for the emergence of
secondary structures, we generated ensembles of random walks and self-avoiding
random walks and compared their helix-content to that of Hamiltonian walks.
Random walks were generated simply from a series of random steps on the square lattice.
Self-avoiding walks were sampled using a Monte-Carlo process based on the pivot algorithm \cite{pivot}.
The results of these computations are shown in
Fig.~\ref{cprwsaw}.

As might be expected, based on the results stated above, the measured helix content
is self averaging (its distribution becomes narrower with increasing $N$), for all three polymer models.
This is explicitly seen in the insets in Fig.~\ref{cprwsaw} where we plot the variance of the
fractional helix content distribution.
We find that there
is a clear difference in the average helical content of random walks and self-avoiding walks
compared to Hamiltonian walks. The three different polymer models have 8\%, 11\%, and 28\% helical content,
respectively, in the long chain limit.

\section{Conclusion}

In this paper we describe and test a Monte-Carlo algorithm for sampling compact polymers on the
square lattice. The algorithm is based on the ``backbite'' move introduced by Mansfield  \cite{Mansfield82} for the
purpose of simulating a many-chain polymer melt. We demonstrate that the algorithm
satisfies detailed balance which ensures that all the accessible states are sampled with the
correct weight. While we have been unable to prove the ergodicity of the algorithm for
large lattice sizes a number of numerical tests seem to indicate its validity. Furthermore,
we measure the efficacy of the algorithm and find that the computational effort (measured
in Monte-Carlo steps per site) grows slightly faster than linear with the polymer length. In practice,
using a pentium-based workstation, it takes roughly an hour  to sample 10000 statistically independent
compact polymer configurations for a chain 2500 monomers in length.

We employ this algorithm in studies of secondary structure of compact polymers on the
square lattice, in the long chain limit. Our results complement the results found previously
for short chains by Chan and Dill \cite{chan89}. Namely, we show that the fraction of
residues participating in secondary structure has a well defined long-chain limit that is
strictly less than one. Looking at helix content alone, we find that helices are twice
as likely to appear in long compact chains then in random walks or self-avoiding walks.
In the context of real proteins this result suggests that hydrophobic collapse to a
compact native state might in large part be responsible for the observed preponderance
of  secondary structures.

The Monte-Carlo algorithm described here for two-dimensional compact polymers
can be easily extended to three dimensions, and various kinds of interactions between
the monomers can be introduced. This will amount to assigning different energies to different compact
chains for which a Metropolis-type algorithm with the backbite move can be employed.
How well the algorithm performs in these situations remains to be seen.

\bibliographystyle{apsrev}

\end{document}